\documentclass[draft]{agujournal2019}
\usepackage{url} 
\usepackage{soul}
\usepackage{subcaption}
\usepackage{amssymb}
\usepackage{amsmath}
\usepackage{amsthm}
\usepackage{mathrsfs}
\usepackage{float}
\usepackage{tikz-cd}
\usepackage{algorithm}
\usepackage{algpseudocode}
\usepackage[normalem]{ulem}
\draftfalse

\journalname{JGR: Machine Learning and Computation}

\newcommand{\D}[0]{\mathcal{D}}
\newcommand{\R}[0]{\mathbb{R}}
\DeclareMathOperator*{\argmin}{arg\,min}

\newcommand{\av}[1]{\textcolor{black}{#1}}

\newcommand{\VL}[1]{\textcolor{black}{#1}}

\newcommand{\acronym}[0]{\texttt{Strass-PINN}\;}

\begin{document}

%
%


\title{Deep learning in the abyss: a stratified Physics Informed Neural Network for data assimilation}

%
%




\authors{Vadim Limousin\affil{1},  Nelly Pustelnik\affil{1},  Bruno Deremble\affil{2} and Antoine Venaille\affil{1}.}


\affiliation{1}{CNRS, ENS de Lyon, LPENSL, UMR5672, 69342, Lyon cedex 07, France}
\affiliation{2}{Université Grenoble Alpes, CNRS, INRAE, IRD,
Grenoble-INP, Institut des Géosciences de l’Environnement, Grenoble, France}




\correspondingauthor{Vadim Limousin}{vadim.limousin@lmd.ipsl.fr}
\correspondingauthor{Nelly Pustelnik}{nelly.pustelnik@ens-lyon.fr}
\correspondingauthor{Bruno Deremble}{bruno.deremble@univ-grenoble-alpes.fr}
\correspondingauthor{Antoine Venaille}{antoine.venaille@ens-lyon.fr}



\begin{keypoints}
\item We propose a multi-layer Physics Informed Neural Network (PINN) suited for deep flow reconstruction. 
\item The method is illustrated  using SWOT-like and ARGO-like pseudo-observations in a three layer quasi-geostrophic model of midlatitude oceans.
\end{keypoints}

%
%

%
%

\justifying
\begin{abstract}
The reconstruction of deep ocean currents remains a major challenge in data assimilation due to the scarcity of interior observations. In this work, we present a proof of concept for deep ocean flow reconstruction using a Physics-Informed Neural Network (PINN). We introduce an algorithm called \acronym (for Stratified Assimilation with  PINNs), which assigns a separate network to each layer of the ocean model while allowing them to interact during training. The neural network takes spatiotemporal coordinates as input and predicts the stream function at those points. Using a SIREN architecture (a multilayer perceptron with sine activation functions), which has proven effective in various contexts, the network is trained using both available observational data and dynamical priors enforced at several collocation points. We apply this method to pseudo-observed ocean data generated from a 3-layer quasi-geostrophic model, where the pseudo-observations include surface-level data akin to SWOT observations of sea surface height, interior data similar to ARGO floats, and a limited number of deep ARGO-like measurements in the lower layers. Our approach reconstructs ocean flows in both the interior and surface layers, demonstrating a strong ability to resolve key ocean mesoscale features, including vortex rings, eastward jets associated with potential vorticity fronts, and smoother Rossby waves. This work serves as a prelude to applying \acronym to real-world observational data and to develop a more comprehensive data assimilation framework using PINNs in ocean context.
\end{abstract}

\section*{Plain Language Summary}

Although bottom ocean currents are crucial for understanding overall ocean circulation, observations in the deep ocean are limited, relying mainly on a few temperature, salinity, and pressure measurements. In contrast, the ocean surface is monitored in high detail by satellites. This article proposes using both the dynamical equations that describe ocean currents and advanced machine learning methods to reconstruct the full ocean dynamics. Since surface and bottom ocean dynamics are correlated, combining physical constraints with statistical analysis can help infer the entire system dynamics. This approach, called Physics-Informed Neural Networks (PINNs), integrates physics and statistical tools into a unified learning architecture.

%
%

%


%
%
%
%

\section{Introduction}

While satellites provide a vast amount of data at the ocean surface \cite{fu1994topex, fu.pavelsky.ea_2024}, the interior remains sparsely observed, due to extreme conditions. Instruments such as  
localized moorings, ARGO floats \cite{wong2020argo}, and acoustic measurements \cite{Munk_Worcester_Wunsch_1995} offer valuable insights into deep layers, but reconstructing deep ocean dynamics from such limited data remains a major challenge. Reconstructing deep ocean currents
is of tremendous importance, as deep flows play a central role in heat storage, carbon sequestration, and the oceanic energy budget \cite{fox2021ocean}.

From a methodological point of view, the uneven distribution of data makes this inverse problem particularly challenging. Traditionally, this issue has been addressed using either statistical methods, e.g. \cite{guinehut2012high}, which are agnostic to the underlying flow model, or data assimilation techniques, which combine available data with a flow model to provide reanalysis \cite{blayo2014advanced, carrassi2018data}. Over the past decade, there has been a growing shift toward machine learning techniques \cite{sonnewald2021bridging}. 

\av{In this paper, we focus on Physics-Informed Neural Networks (PINNs) \cite{karniadakis2021physics}, a versatile framework that can solve both forward problems (e.g., solutions of partial differential equations without observations) and inverse problems (e.g., integrating observations with physical constraints). In the latter context, PINNs acts as a machine learning counterpart to data assimilation, combining observational data with a flow model to reconstruct the system state \cite{blayo2014advanced}. However, when measurements are sufficiently dense, the method operates closer to interpolation, with the flow model acting as a regularization term. This is the regime we explore here, where we leverage PINNs for ocean flow reconstruction from abundant surface data and sparser subsurface data.}

\textcolor{black}{A key feature of PINNs is that the field to be recovered is encoded directly in the neural network, which can reduce the reliance on grid-based discretization of the field. However, before this potential advantage can be meaningfully compared against other methods, a major challenge remains: establishing how to train the network to reconstruct multiscale, chaotic oceanic flows. Here, we develop a multilayer PINN learning strategy called \acronym (for Stratified Assimilation with PINNs), and demonstrate its feasibility  for deep flow reconstruction, achieved in the presence of abundant surface data and sparser subsurface data.}

A large body of studies have been focused on reconstructing surface and subsurface fields from these data. For instance, regarding sea surface height (SSH) fields obtained from satellite altimeters, existing gridded datasets are produced using optimal interpolation \cite{ballarotta2019resolutions}. Spatial resolution of the reconstructions can be improved using simple assimilation methods, such as nudging with simple and fast dynamical models \cite{leguillou.metref.ea_2021}, or deep learning strategies with convolutional networks \cite{siegelman2021deep}. More recently, powerful data-driven techniques such as 4DVarNet have replaced the dynamical constraints in the assimilation variational problems with a penalty term involving a neural network in order to learn the dynamic from data themselves 
\cite{beauchamp20234dvarnet}. Other data-driven methods, decompositions into empirical orthogonal functions or autoencoders \cite{barth2020dincae}, have been used to reconstruct sea surface temperature (SST) fields in the presence of clouds that leads to spatio temporal gaps in the data set. Learning efficient interpolation from both SST and SSH has also been addressed within various machine learning frameworks, e.g.  \cite{martin2023synthesizing,fablet2024inversion,archambault2024}. Another line of recent work used vision transformers to learn from idealized simulations the operator that relate surface buoyancy field to interior flow \cite{shan2026testing}.

The common feature of all the techniques mentioned above is the reconstruction of a 2d+t field. Data-driven methods have also been developed for 3d+t fields, provided there is sufficient data for at least some period of time. For example, \citeA{guinehut2012high} used optimal interpolation and multilinear regression on climatological data to infer the vertical structure of mesoscale turbulence from observations alone. Similarly, \citeA{oulhen2024reconstructing} combined data reduction techniques based on Empirical Orthogonal Functions (EOF) and analogs to reconstruct 3d+t temperature fields in the ocean's upper layers, leveraging the ARGO dataset to learn statistical properties and reconstruct past fields from before the ARGO era.  While promising, these methods are not ideal for reconstructing fields in regions that have never been intensively sampled, as statistical inference may not be reliable in such areas. For instance, trapped flow over seamounts may not always exhibit a clear signature at the surface \cite{venaille2011baroclinic,deremble2023eddy}. While  moorings or deep ARGO floats provide useful local information on deep flows, global maps of deep ocean circulation can only be  inferred using data assimilation techniques that combine data with flow models  \cite{flierl1978models}.
 This approach has provided \av{invaluable} reanalysis datasets, even if some unexpected behavior is also reported in some regions, see e.g. \citeA{jean2021copernicus}. While successful for operational purposes, assimilation methods such as 4DVAR or the Ensemble Kalman Filter are often difficult to implement and computationally expensive. On the other hand, SSH interpolation via deep learning have been proven skillfull to reconstructing subsurface mesoscale structures \cite{siegelman2021deep}. In all those cases, combining data from observations with dynamical constraints or deep learning approaches  typically requires the introduction of a spatial grid, which incurs a significant computational cost as the spatio-temporal resolution increases. This highlights the need for more flexible methods that can incorporate dynamical constraints within a mesh-free framework. This is precisely where PINNs could offer potentially an advantage.

In a PINN, the physical state of the system (such as current velocity and density) is represented as the output of a neural network, while the input consists solely of spatio-temporal variables. The specificity of PINNs training is to ensure that the solution generated by the trained neural network not only satisfies the observed data but also adheres to the differential equations that describe the system's dynamics. This is achieved by designing a training loss function composed of a data fidelity term, which ensures consistency with the observed data, and a penalty term, which quantifies the discrepancy between the network output and the assumed physical dynamics. The use of PINNs provides a threefold benefit: it bypasses the need to specify artificial boundary conditions when constructing the field in a given region, it is mesh free,  and it provides useful flexibility in the choice of flow models, which can be adapted on a case-by-case basis.

We use here the term ``meshless" in the modern machine learning context to refer to methods that do not depend on mesh connectivity or basis functions defined over mesh elements. However, the fields are still represented using a finite number of parameters and are evaluated at observation and collocation points where physical constraints are enforced during training. The advantage of such implicit neural representations, compared to gridded fields, is their ability to compress data while retaining useful properties, such as differentiability at any point in the field or the ability to represent complex geometries. These benefits have been quantitatively demonstrated in diverse applications, including image analysis and the encoding of atmospheric data \cite{dupont2021coin,dupont2022data}.

Following the original proposal of \citeA{raissi2019}, PINNs have  been  applied to a wide range of fluid mechanics problems, from Rayleigh-Bénard convection \cite{lucor2022simple, clark2023reconstructing} to denoising measurements in stratified turbulence \cite{zhu2024new}, among a vast number of other examples \cite{karniadakis2021physics}. Yet, applications  of PINNs in atmospheric or oceanic field reconstruction are still in the early stages, see e.g.  \citeA{millet2023physics,yoon2024predicting}. Notably, \citeA{bang2024physics} addressed recently the ability of PINNs to reconstruct surface velocity field from surface drifter data. \textcolor{black}{Implementing the PINN method in a realistic context such as mesoscale ocean turbulence (from $30$ to $300$km) presents several challenges, especially when multiple temporal and spatial scale are involved. A first challenge, commonly encountered in deep learning, is to prevent spectral bias effects, i.e., the inability to learn the high frequencies of a signal over a reasonable training time \cite{rahaman2019spectral}. A second challenge, more specific to PINNs, is that training these networks involves a delicate balance between ensuring data fidelity and adhering to physical laws, which can complicate the optimization process and lead to convergence issues, requiring the use of specialized techniques and strategies \cite{karniadakis2021physics,rathore2024challenges}.}

\textcolor{black}{In the present work, we show how to adapt PINNs methodology to reconstruct deep ocean flows by leveraging surface observations and sparse data from deeper layers. Compared to state-of-the-art PINNs learning techniques, the proposed \acronym approach introduces a novel learning strategy by incorporating a multilayer training specifically tailored to multilayered ocean models.  We demonstrate that the proposed PINNs can propagate information between vertical layers, reconstructing three-dimensional flow dynamics even with limited direct measurements from the deep ocean layer.}

\begin{figure}[htbp!]
    \centering
    \includegraphics[width=\linewidth]{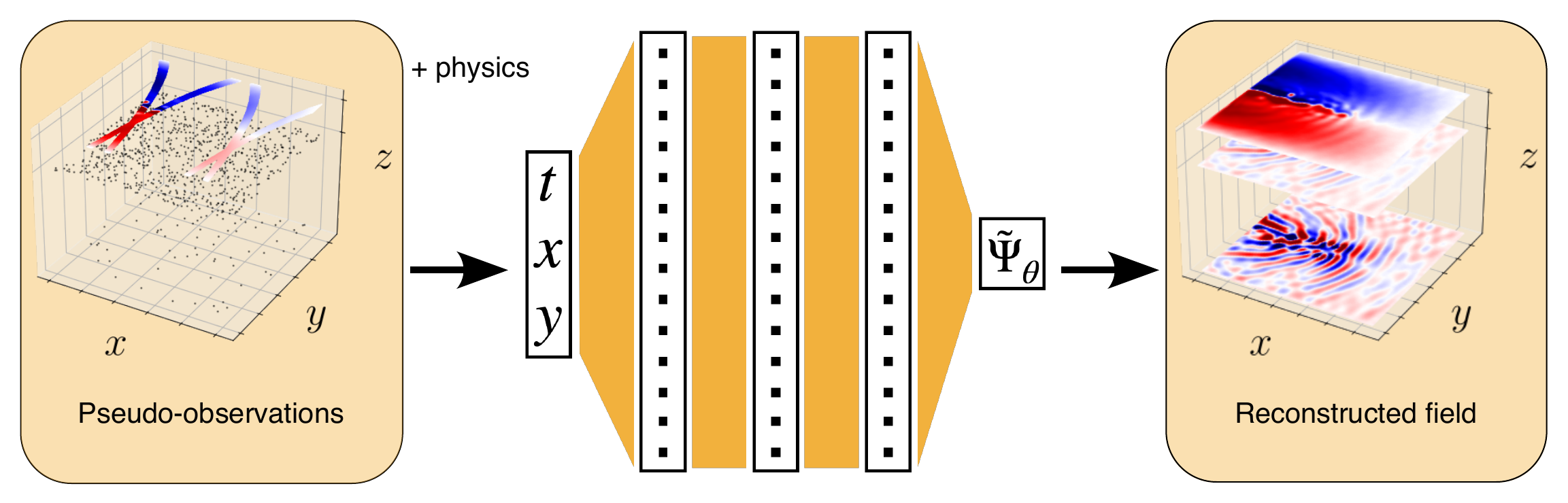}
    \caption{\av{Pseudo-observations are generated through satellite-like sampling of the surface flow and buoy-like sampling of the interior layers in a three-layer quasi-geostrophic model. The goal of \acronym is to reconstruct the full model flow, including regions not directly observed, by incorporating physical constraints from the dynamical equations of the ocean model. The reconstructed flow is represented by a neural network that takes spatio-temporal coordinates as input and outputs the stream function. Training involves minimizing a loss function that combines a data fidelity term (to match the observations) and a dynamical residual term (to enforce consistency with the governing equations). Once trained, the network can be evaluated at any point in space and time. In practice, reconstructed fields are computed on a grid spanning the entire domain for comparison with the ground-truth simulation.}}
    \label{method_sketch}
\end{figure}

\textcolor{black}{For that purpose we consider a simplified wind-driven three-layer ocean model, a minimal yet representative framework for ocean dynamics. The upper layer is observed through surface data (similar to SWOT-like altimetry observations), while the interior and deep layers are observed by sparse point measurements (similar to ARGO floats), as sketched in Fig.~\ref{method_sketch}. By integrating these observations with the known system dynamics, we show that is it possible to train PINNs in order to reconstruct deep ocean flow, in semi-realistic settings. This work serves as a proof of concept for the use of PINNs in ocean flow reconstruction, particularly in regions where data are sparse or indirect. This is a necessary first step before real-world oceanographic applications, including uncertainty quantification and comparison with other existing methods, to be addressed in future work.}

\section{State of the art and context}
\subsection{Data Assimilation}

Data assimilation is an inverse problem where one recovers a signal from partial or degraded measurements, in our case partial sea surface data from satellites and sparse subsurface data from buoys. The typical approach to assimilation is a grid-based approach that can be framed as a minimization problem, where the cost function consists of two terms: a data fidelity term and a penalty that enforces predefined dynamics. The solution to the assimilation problem corresponds to the minimiser of this cost function. Formally, the goal is to recover the temporal sequence of 3d states $\mathrm{u}(t,x,y,z)$, with each component of $\mathrm{u}$ representing different variables (e.g., pressure, velocity, temperature), and $(t,x,y,z)$ indicating spatio-temporal locations on a grid of size $N = N_t \times N_x \times N_y \times N_z$. Assuming partial observations of the system at $M$ distinct spatio-temporal points (represented by the vector $\mathrm{d}$ of size $M$), the cost function to be minimized is defined as:
\begin{linenomath*}\begin{equation}
    \mathcal{L}_{\texttt{DA}}(\mathrm{u}) =  \Vert \mathcal{H}( \mathrm{u})  - \mathrm{d} \Vert + \|\mathcal{D}(\mathrm{u})\| 
\label{DA}
\end{equation}\end{linenomath*}
where $\mathcal{H}$ represents the observation operator, which establishes the relationship between the complete quantity to be recovered of size $N$ and the grid locations where observations are available. Additionally, $\mathcal{D}$ represents the physical model. The first term is referred to as the data term and ensures the reconstructed field matches the observed data $\mathrm{d}$ as well as possible. 
The second term is the dynamic information and ensures the reconstructed time series follows physical laws. 
Typically, $\mathcal{D}$ is a discretisation of the explicit dynamics, formulated as a partial differential equation on the $N_t\times N_x\times N_y\times N_z$ grid, and the norm $\|\,\cdot\,\|$ can be either a squared $L^2$, $L^1$, $TV$, or Mahalanobis distance.

In this procedure, the physical model can represent multiple scales processes and the major hurdle is then to compute the adjoint operator of this model to minimize (\ref{DA}). This minimization problem is the one behind the 4DVar methods \cite{bocquet2014}. While 3DVar assimilates observations at a single time point, 4DVar extends this over a time window to better capture system dynamics, at increased computational cost. Kalman filtering or extended version for non-linear systems such as Extended Kalman Filter (EKF) or Ensemble Kalman Filter (EnKF), on the other hand, adopt a probabilistic and sequential approach, updating the state estimate at each time step while dynamically propagating uncertainties, but remains computationally expensive, especially for non-linear or large-scale systems.

Several attempts to include machine learning in these approaches have already been explored (e.g., 4DVarNet) \cite{deepDA, Fablet_2021, beauchamp20234dvarnet}, combining standard assimilation techniques with neural networks either end-to-end or to improve the modeling of $\mathcal{D}$. More precisely, the network predicts $\mathrm{u}_{t+\Delta t}$ given $u_t$ and the regularisation term involves $\mathcal{D}(\mathrm{u}) = \mathrm{u}_{t+\Delta t} - f_\theta(\mathrm{u}_t)$, encouraging the reconstructed field $\mathrm{u}$ to adhere to the learned dynamics modeled by the neural network denoted $f_\theta$, with $\theta$ the network parameters. 
They show very promising results already comparable to classical approaches. For instance \citeA{beauchamp20234dvarnet} has reported a ``\textit{relative improvement with respect to the operational optimal interpolation between 30\% and 60\% in terms of reconstruction error}''. However, these methods are often limited to surface reconstruction and do not address the ocean interior, while also being tied to a fixed grid whose resolution directly impacts training cost.




\subsection{PINNs}

A second class of approaches relies on Physics-Informed Neural Networks (PINNs)
which is at the core of this contribution.   PINNs share the formulation \eqref{DA}, with the key distinction that $\mathrm{u}$ is modeled by a neural network $f_\theta(t,\mathrm{x})$ taking spatio-temporal coordinates as inputs. This implicit neural representation of fields (also called neural fields) parameterized by the network weight and bias contrast with more traditional representations on grid, or decomposed on spectral components. Let $\mathrm{u}_\theta(t,\mathrm{x})$ be a neural network parameterized by $\theta$ approximating the desired field at any spatio-temporal location. The PINN formulation seeks the optimal $\theta$ minimizing:
\begin{linenomath*}
\begin{equation}
    \mathcal{L}_{\texttt{PINN}}(\theta) = \frac{1}{n} \sum_{(t,\textrm{x}) \in \mathcal{I}} \left( \mathrm{u}_\theta(t,\textrm{x}) - \mathrm{d}(t,\textrm{x}) \right)^2 + \lambda \| \mathcal{D}(\mathrm{u}_\theta, \partial_t \mathrm{u}_\theta, \nabla \mathrm{u}_\theta, \dots) \|^2.
\label{obj}
\end{equation}
\end{linenomath*}
where $ \mathcal{I} $ denotes either a set of $n$ points where measurements $\mathrm{d}$ are available (for data assimilation problems), a set of $n$ points that sample initial and boundary conditions (to solve a well-posed Cauchy problem), or a combination of both (PINNs were originally developed to solve well-posed Cauchy problems with prescribed initial and boundary conditions and known dynamics). 

The loss function thus consists of a data term and a regularization term measuring deviation from the dynamical prior. It serves as an objective function that encourages the minimiser to both closely solve the prior PDE $ \mathcal{D}(\mathrm{u}, \partial_t \mathrm{u}, \nabla \mathrm{u}, \dots) = 0 $ and ensure fidelity to the observations $\mathrm{d}$. The norm $ \| \mathcal{D}(\cdot) \| $ represents an $ L^2 $ norm over the spatio-temporal domain of interest. \av{In practice, this integral is replaced by a finite sum over a set of collocation points randomly sampled from the domain  (following classical Monte Carlo method). Importantly, this involves evaluating the residual at a discrete set of collocation points that do not necessarily coincide with the data points.}

The computation of the loss function assumes that the network $ \textrm{u}_\theta $ can be differentiated with respect to its input coordinates to compute $ \partial_t \textrm{u}_\theta$, $ \partial_x \textrm{u}_\theta $, $ \partial_y \textrm{u}_\theta $, and any higher-order derivatives required by the prior PDE. In practice, this step is straightforward due to \textcolor{black}{automatic differentiation \cite{baydin2018automatic}}, which allows handling a wide range of differentiable functions $ \textrm{u}_\theta $. This is a major advantage of PINNs formulation. This approach is considered mesh-free because the field $ \textrm{u} $ is not discretized on a grid, unlike standard assimilation techniques.

\textcolor{black}{One the one hand, the interest of using implicit neural representation in ocean context has been shown by \cite{johnson2022neural}, yet without considering additional physical constraints. On the other hand, the potential of PINNs for inverse problems such as data assimilation in fluids has long been realized \cite{raissi2020hidden}. Our aim here is to bridge these two studies, by adapting the PINNs approach to a configuration of oceanic relevance.}

In its original formulation, PINNs were implemented using a multilayer perceptron with a tanh activation function for $\textrm{u}_\theta$, alongside a Latin Hypercube Sampling strategy to select collocation points where the PDE residual is computed \cite{raissi2019}. Since then, significant efforts have focused on optimizing the choice of collocation points \cite{nguyen2022fixed}, exploring alternative network architectures \cite{wang2024kolmogorov}, and testing different activation functions \cite{sitzmann2020implicit}, among other strategies \cite{karniadakis2021physics}. 
The algorithmic strategy itself is rarely studied, as it generally relies on a single neural network fully trained using standard optimization tools. In this study, the specific layered structure of the underlying ocean dynamics will lead us to focus on both the choice of the architecture and the algorithmic strategy.




\subsection{Focus on the implicit neural representation SIREN}

We use the SIREN architecture, a fully connected network with sinusoidal activations and a tailored initialization \cite{sitzmann2020implicit}, chosen for its ability to simultaneously fit a function and its gradients. This network architecture has proven useful in reconstructing various natural signals, including flow reconstruction with PINNs in the context of Rayleigh-Benard convection \cite{mommert2024periodically}. SIREN networks have also been used in ocean context, albeit without physical constraints, to represent surface current as a neural field \cite{johnson2022neural}.

\av{In this study, we consider SIREN networks, with $M+1$ layers and $w$ neurons in each hidden layer, to describe $2d+t$ scalar fields:}
\begin{equation}
\label{SIREN}
    \Psi_\theta(t, x, y) = W_M\cdot\sin(\ldots W_2\cdot\sin(W_1\cdot\sin(K_xx + K_yy - \Omega_tt + \varphi) + b_1) + b_2) \ldots + b_M
\end{equation}
\noindent \av{where the sine activation function is applied component-wise to its vector input.} 
Network parameters are $$\theta = \{K_x, K_y, \Omega_t,\varphi,W_1,\ldots, W_M, b_1,\ldots, b_M\},  $$ where $K_x, K_y, \Omega_t,\varphi, \in \R^{w}$, $ b_m \in \R^{w}$, $W_m \in \R^{w\times w}$, $W_M \in \R^{1 \times w}$, and $b_M \in \R$.

This implicit neural representation takes field coordinates as inputs and outputs a scalar estimate of the field.  The notion of resolution then shifts from a spatial resolution on a discretised domain to an implicit resolution related to the size of $\theta$. Empirically, this implicit resolution corresponds to the spectral resolution. The smallest spatial and temporal scales that the SIREN can generate are limited by the highest values of $K_x, K_y$ and $\Omega_t$.  Therefore, these values must be carefully initialized to ensure the network can capture the small physical scales in the signal.

SIREN is particularly well-suited for our task since its sinusoidal activations are stable under differentiation: derivatives of a SIREN remain a SIREN with shifted weights, including spatial derivatives and higher-order operators such as the Laplacian. This property explains why SIREN can simultaneously fit a function and its derivatives \cite{sitzmann2020implicit}, making it ideal for incorporating PDE-based physical constraints. Originally introduced for image-related problems and briefly tested as a PINN on a simple wave equation, we here exploit its full potential for a more complex ocean assimilation task.

\section{\acronym: new physics informed neural network for data assimilation using layered ocean models}

The proposed \acronym method builds on PINNs by introducing a multilayer structure to both the training loss and the training process. In this section, we present the method through a concrete reconstruction problem for wind-driven, multilayer quasi-geostrophic dynamics, using ground truth data from a simulation of a classic double-gyre setup. This setup captures key features of midlatitude oceans, such as the Gulf Stream and the Kuroshio. Pseudo-observations will be extracted from this quasi-geostrophic ground truth (GT), and the reconstructed flow will be compared to the full simulation field. For clarity, the method is presented here for a three-layer quasi-geostrophic model, though it can be extended to more layers \av{and could be adapted in future work to multilayer shallow water models. Details of the general multilayer formulation are provided in the \ref{appen-sec:general}.}

\av{We emphasize that there are two distinct notions of "layers" in this work: oceanic layers and neural network layers. We will clarify the context to avoid confusion whenever necessary. Additionally, we stress that ocean data assimilation with PINNs does not require layered models. It could instead rely on a single network describing a 3D+t field, with physical constraints derived from 3D primitive equations. Our hypothesis here is that the layered formulation may facilitate the transfer of information from the surface (where abundant data is available) to deeper layers during network training.}

\subsection{Dynamical model: three-layer quasi-geostrophic dynamic}
\label{MultiQG}

Due to stable vertical density stratification and the small vertical-to-horizontal aspect ratio, ocean models are often represented as layers stacked on top of one another. In these models, stratification is defined by the vertical gradient of density. \av{At midlatitudes, where the Coriolis force dominates over inertial effects (i.e., the Rossby number is small), the dynamics of large-scale flows can be effectively described by the quasi-geostrophic  model \cite{vallis2017atmospheric}. This model is derived from an asymptotic expansion of the shallow water equations at low Rossby number, filtering-out fast inertia gravity waves. The resulting set of equations is simpler that the original shallow water model, as it involves less prognostic fields.} 
We focus on a three-layer quasi-geostrophic model driven by wind, which filters out fast motions while capturing key mid-latitude mesoscale features: western boundary current intensification, Rossby wave propagation, eastward jet formation, and mesoscale vortices. Three layers suffice to reproduce these key features of oceanic turbulence, providing a minimal framework for understanding currents such as the Gulf Stream or Kuroshio \cite{miller.deremble.ea_2025}.

\begin{figure}[htbp!]
\begin{tabular}{cc}
\includegraphics[height=4cm]{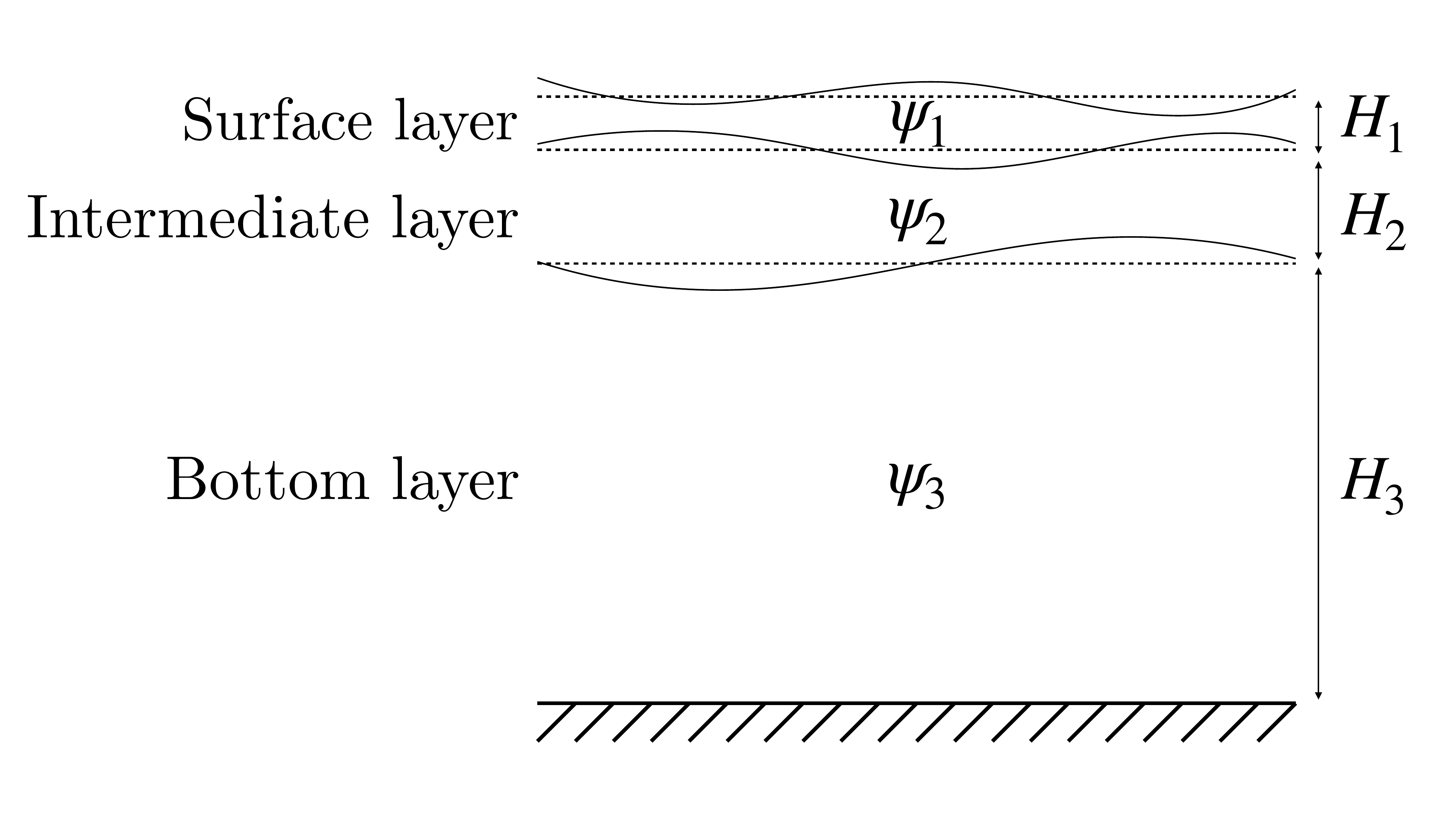}&
        \includegraphics[height=4cm]{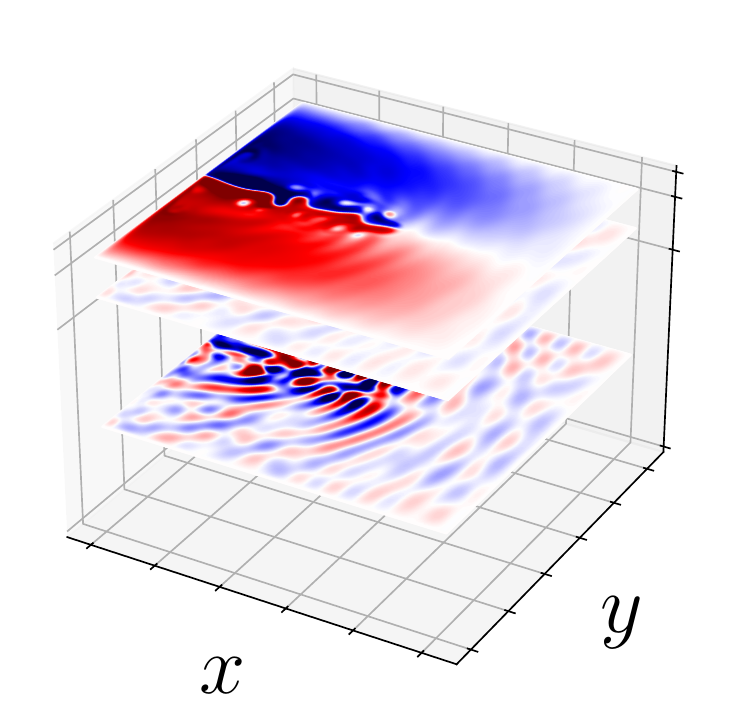}\\
     (a) Sketch of the model &
     (b) Typical output of the model
     \end{tabular}    
    \caption{(a) Illustration of the three-layer framework with 2d flows in each layer encoded in stream functions $\psi_i$. (b) Maps of the stream functions in a three layers quasi-geostrophic model of the ocean. Three important features appear in the generated stream functions. First, as expected with the choice of wind forcing, two western intensified gyres are formed at the surface and induce a jet stream in the center of the domain.  Second, we notice Rossby waves propagating in each layer. \av{Third, ocean vortex rings are shed on the side of the jet stream and propagate upstream, towards the west.}}
    \label{qg_framework}
\end{figure}

We consider three  layers of homogeneous density $(\rho_1,\rho_2,\rho_3)$ stacked vertically, as shown in Fig.~\ref{qg_framework}(a).
In each layer, the flow is 2D, non-divergent, and can be described by stream functions $(\psi_1,\psi_2,\psi_3)$, which fully characterize the dynamics as scalar functions of $(t,x,y)$.  The flow takes place in a \av{closed} basin with free slip boundary conditions. The system is forced  by zonal wind stress curl  $ -\partial_y\tau_x(y)$, which drives the upper layer, and dissipates through the Ekman layer in the deeper layer, or via horizontal viscous dissipation in each layer. To capture key features of midlatitude ocean dynamics, we need to incorporate latitudinal variations in the Coriolis parameter using the $\beta$-plane approximation, which assumes linear variations of planetary vorticity with latitude. The resulting system of equations is expressed as the following set of partial differential equations:
\begin{linenomath*}\begin{equation}
\begin{aligned}
    \left.\frac{d}{dt}\right|_1 \underbrace{\left[ \Delta \psi_1 - f_0^2\frac{\psi_1 - \psi_2}{H_1g'_1} + \beta y \right]}_{= q_1}   & = -\frac1{H_1}\partial_y\tau_x + \nu \Delta q_1\\
    \left.\frac{d}{dt}\right|_2 \underbrace{\left[ \Delta \psi_2 - f_0^2\frac{\psi_2 - \psi_1}{H_2g'_1} - f_0^2\frac{\psi_2 - \psi_3}{H_2g'_2} + \beta y \right]}_{= q_2}  & = \nu \Delta q_2 \\
    \left.\frac{d}{dt}\right|_3 \underbrace{\left[ \Delta \psi_3 - f_0^2\frac{\psi_3 - \psi_2}{H_3g'_2} + \beta y \right]}_{= q_3}   & = -r_E \Delta \psi_3 + \nu \Delta q_3
\end{aligned}
\label{qg}
\end{equation}\end{linenomath*}
with $ \left. \frac{d}{dt} \right|_\ell \equiv \partial_t - (\partial_y\psi_\ell) \partial_x + (\partial_x\psi_\ell )\partial_y$ representing the Lagrangian derivative, $ g'_\ell = g \cdot (\rho_{\ell+1} - \rho_\ell) / \rho_\ell $ is the reduced gravity between the $\ell$-th and $(\ell+1)$-th layers, and $ H_\ell $ denotes the height of the $\ell$-th layer. The parameter $ r_E = f_0 h_E / 2 H_3 $ is the bottom friction  coefficient, with $ f_0 $ the Coriolis parameter and $ h_E $ the Ekman layer depth. The double-gyre framework amounts to choosing $ \tau_x = \tau_0 \cos(2\pi y / L) $, with $ L $ as the domain size and $ \tau_0 $ the maximum wind stress amplitude. Additionally, we made the classical rigid \av{lid approximation}.
Typical flow fields obtained with this model are shown in Fig.~\ref{qg_framework}(b) for parameters presented in Tab.~\ref{tab:coeffs}.\\

 \av{Our parameter choices are based on established quasi-geostrophic modeling studies and represent typical mid-latitude ocean conditions, with some necessary compromises for numerical tractability, see e.g. \cite{hogg.killworth.ea_2005}.}

\av{The reduced gravity values ($g'_1 = 0.025$~m\,s$^{-2}$, $g'_2 = 0.0125$~m\,s$^{-2}$) are chosen to yield realistic deformation radii of 39~km and 22~km for the first and second baroclinic modes, respectively \cite{berloff.hogg.ea_2007}. These values are consistent with observations of mid-latitude deformation radii \cite{chelton.deszoeke.ea_1998}. The layer thicknesses ($H_1 = 350$~m, $H_2 = 750$~m, $H_3 = 2900$~m) represent the typical thermocline structure \cite{hogg.dewar.ea_2003}. The values $f_0 = 9.4 \times 10^{-5}$~s$^{-1}$ and $\beta = 1.7 \times 10^{-11}$~m$^{-1}$\,s$^{-1}$ correspond to 40$^\circ$N latitude, calculated as $f_0 = 2\Omega \sin(\phi)$ and $\beta = 2\Omega \cos(\phi)/a$, where $\Omega = 2\pi/86400$~s$^{-1}$, $\phi = 40^\circ$N, and $a = 6.4 \times 10^6$~m is Earth's radius.}

The wind stress amplitude ($\tau_0 = 1.3 \times 10^{-5}$~m$^2$\,s$^{-2}$, equivalent to $0.01$~N\,m$^{-2}$) is weaker than typical observed values ($\sim 0.1$~N\,m$^{-2}$). This choice, following common practice in quasi-geostrophic modeling studies, ensures the jet penetrates appropriately across the domain (for stronger values of the wind, the jet would collide with the eastern boundary). Alternatively, one could also have used other dissipation mechanisms such as no-slip boundary conditions \cite{miller.deremble.ea_2024, miller.deremble.ea_2025, deremble.hogg.ea_2011}. Similarly, the 4000$\times$4000~km$^2$ domain, while smaller than the full North Atlantic, is consistent with regional quasi-geostrophic studies and allows us to focus on the essential double-gyre dynamics.

For these parameters, the turbulence level is maximum near the jet, with more quiescent conditions in the rest of the gyres. As far as field reconstruction is concerned, this spatial heterogeneity is interesting, as it provides two distinct dynamical regimes within a single domain: a turbulent regime characterized by intense mesoscale activity in the central jet region, and a regime dominated by Rossby wave propagation in the outer gyre areas. While this separation is not as pronounced in the real ocean, it creates a useful test case for our reconstruction methodology, allowing us to evaluate performance across dynamically different flow regimes within the same simulation. This dual-regime structure is particularly relevant for our focus on deep current reconstruction, where the ability to capture both energetic turbulent features and large-scale wave-like motions is crucial for understanding the full spectrum of deep ocean dynamics.

A shortcoming of the quasi-geostrophic model is that it does not resolve submesoscale features such as filaments and small-scale, surface-intensified vortices. Given the known limitations of current PINNs in recovering multiscale flows, we deliberately chose a regime that does not contain these finer-scale motions. This allows us to focus on validating the \acronym framework in a controlled, yet dynamically rich, mesoscale environment.

\begin{table}[h]
    \centering
    \begin{tabular}{|c|c||c|c|}
        \hline
        Coefficient & Value & Coefficient & Value \\
        \hline
        $H_1$ & 350 m & $\beta$ & $1.7\times 10^{-11}$ m$^{-1}\cdot$s$^{-1}$ \\
        $H_2$ & 750 m &  $f_0$ & $9.4\times 10^{-5}$ s$^{-1}$ \\
        $H_3$ & 2900 m & $\tau_0$ & $1.3\times 10^{-5}$ m$^2\cdot$s$^{-2}$ \\
        $g'_1$ & $0.025$ m$\cdot$s$^{-2}$ & $\nu$ & 40 m$^{2}\cdot$s$^{-1}$ \\
        $g'_2$ & $0.0125$ m$\cdot$s$^{-2}$ & $h_E$ & 2 m \\
        $L$ & 4000 km & & \\
        \hline
    \end{tabular}
    \caption{\av{Parameters used in the quasi-geostrophic model setup.}}\vspace{-0.3cm}
    \label{tab:coeffs}
\end{table}


\subsection{Posing the reconstruction problem with pseudo-observations}
\label{Pseud-Obs prob}

Our idealized quasi-geostrophic framework aims to address a typical ocean assimilation problem. In practice, abundant surface data are available, mostly from satellites such as SWOT. Subsurface data, sparser and less frequent, are provided by instruments like Argo floats. In real observations, Argo floats typically measure temperature, salinity, and pressure at a depth of 1~km for 9~days. They then descend to 2~km before returning to the surface to transmit data, completing a 10-day cycle. The global Argo array comprises more than 3000~active floats, which corresponds to about 500~floats in a typical ocean basin such as the one considered in our idealized simulation. A smaller number of Deep Argo floats extend measurements to 6 km, sampling the full water column.

In our three layer model, pseudo-observations will be represented by satellite SWOT-like data for the first layer, sparser Argo-like random measurements for the second layer, and even sparser ARGO-like observations for the third layer.

In a quasi-geostrophic framework, SWOT-like observations of sea surface height provide direct measurements of the upper-layer streamfunction (denoted $\psi_1$ in the model above). The salinity, temperature, and pressure measurements from Argo floats can be combined to obtain a density profile using the standard seawater equation of state. Using hydrostatic balance and geostrophy, Argo-like observations in a layered model provide a measure of the streamfunction difference between adjacent layers, i.e.\ $\psi_{\ell+1} - \psi_\ell$ at level $\ell$.

\begin{figure}[htbp!]
    \centering
    \begin{subfigure}{0.3\textwidth}
        \includegraphics[width=\textwidth]{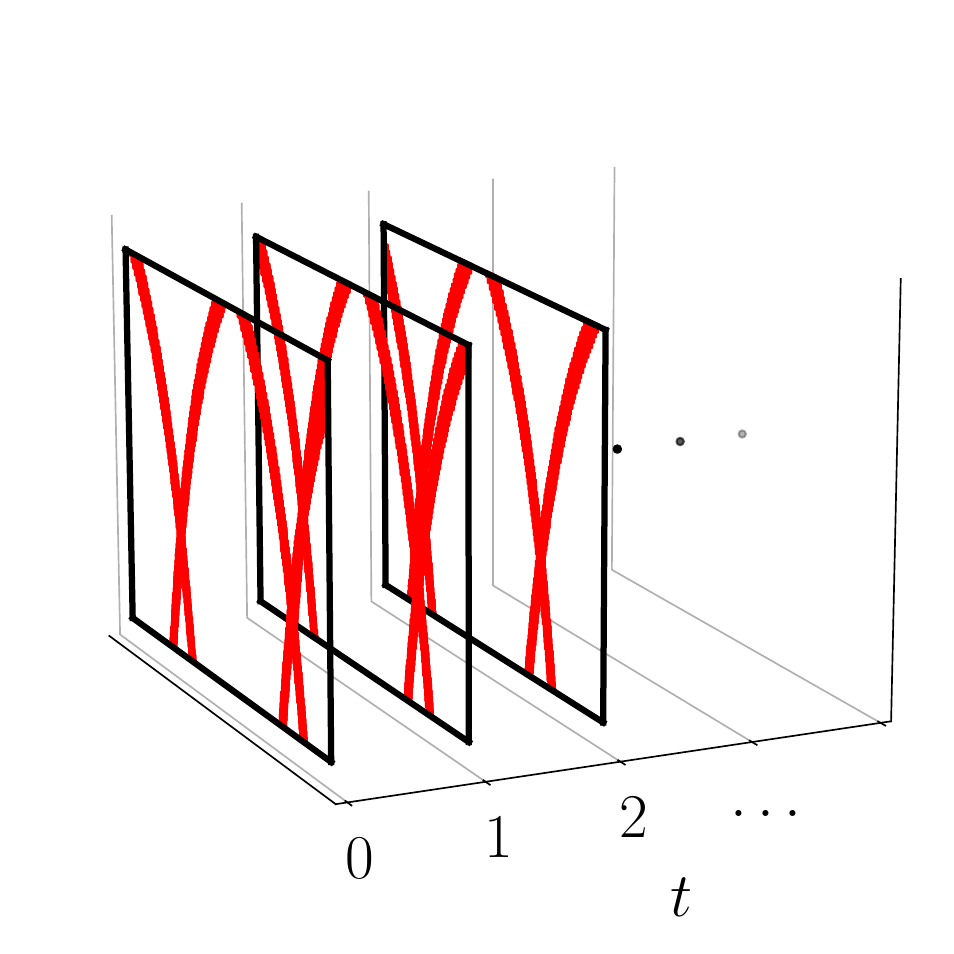}
        \caption*{($\mathcal{I}_1$) SWOT-like.}
    \end{subfigure}
    \hspace{0.1cm}
    \begin{subfigure}{0.3\textwidth}
        \includegraphics[width=\textwidth]{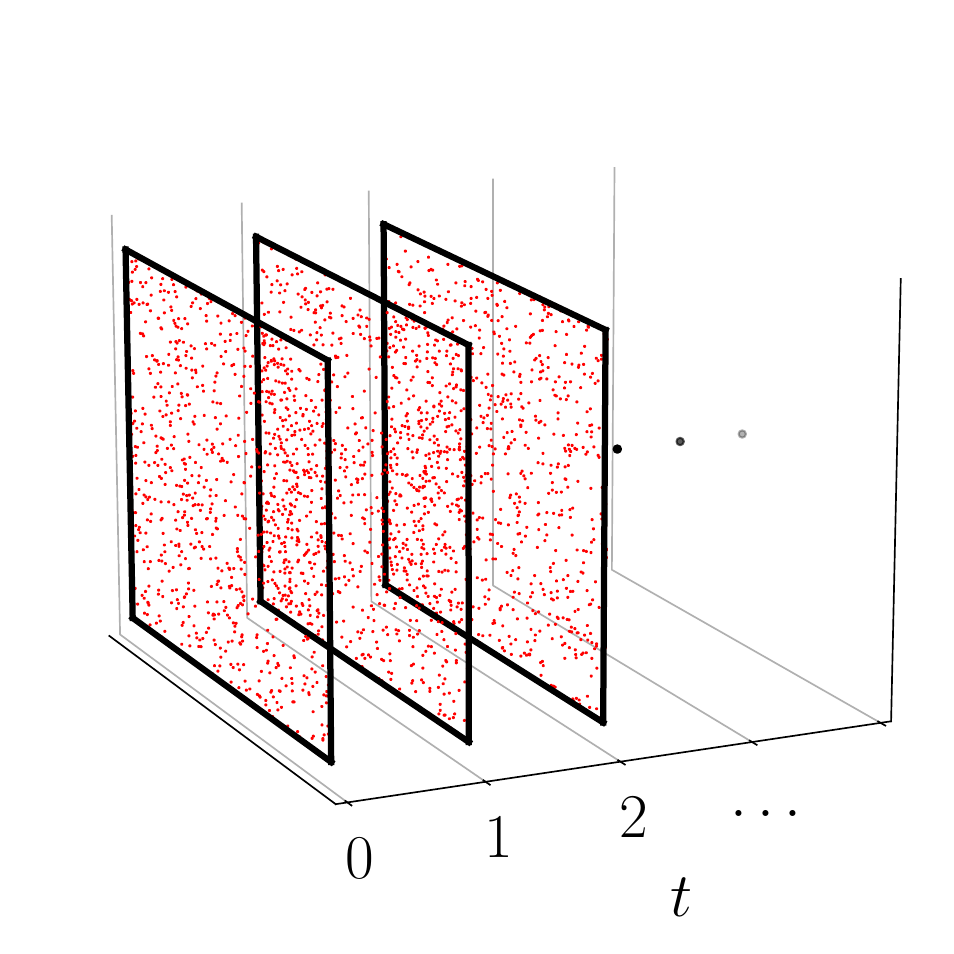}
        \caption*{($\mathcal{I}_2$) Argo-like.}
    \end{subfigure}
    \hspace{0.1cm}
    \begin{subfigure}{0.3\textwidth}
        \includegraphics[width=\textwidth]{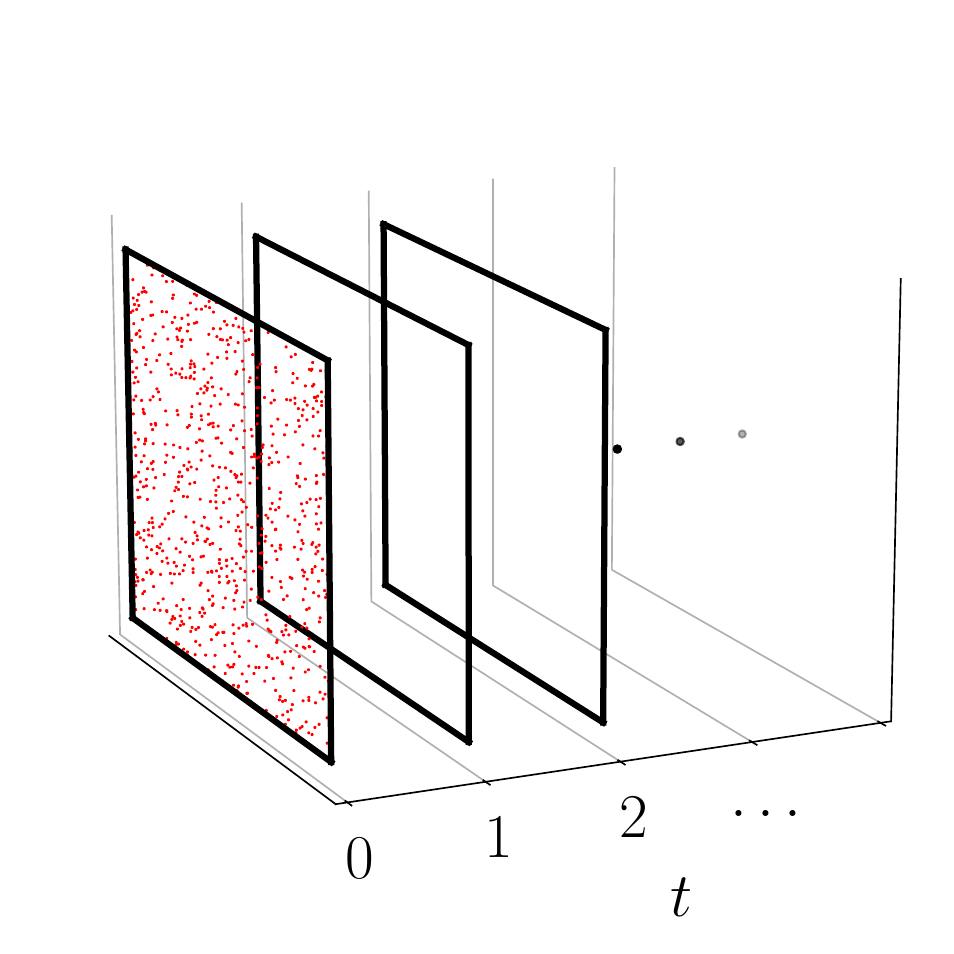}
        \caption*{($\mathcal{I}_3$) Argo-like (deep).}
    \end{subfigure}
    \caption{In our set-up, observations are only provided by pseudo-measurement devices at locations stored in $\mathcal{I}_1$, $\mathcal{I}_2$, and $\mathcal{I}_3$. These three sets are illustrated here. \av{Time unit is day.} Every simulation day, we have access to SWOT-like measurements of $\psi_1$ and Argo-like measurements of $\psi_2 - \psi_1$ and every 10 simulation day we have Argo-like measurements of $\psi_3 - \psi_2$.\vspace{-0.5cm}}
    \label{pseudo-obs}
\end{figure}

Concretely, following the notation in \eqref{obj}, we define a set of observations $\mathcal{I}_\ell$ for each layer $\ell=\{1,2,3\}$. This set incorporates daily SWOT-like measurements of $\psi_1$, daily Argo-like measurements of $\psi_2 - \psi_1$, and Argo-like measurements of $\psi_3 - \psi_2$ every 10 days, as shown in Fig.~\ref{pseudo-obs}. The complete set of observations is denoted as $\mathcal{I} = \mathcal{I}_1 \cup \mathcal{I}_2 \cup \mathcal{I}_3$. 
After 22~days, float locations are randomly reshuffled across the domain to ensure broad spatial coverage over time. This 22-day reshuffling period is chosen to align roughly with the eddy turnover timescale in the quasi-geostrophic model. It also ensures adequate sampling of the domain without explicitly modeling float advection. More details about this set will be provided in subsection~\ref{sub:data}.

We stress that our pseudo-observation sampling differs from the actual Argo dataset. Real Argo floats typically provide data only during the final ascent phase of their cycle, their measurement may also include non geostrophic effects, and the uneven sampling due to passive advection is not accounted for here. While applying this framework to real datasets and more complex models requires further work, our setup captures the essence of real-world ocean observing systems with dense surface and sparse subsurface data, serving as a proof of concept for PINNs in ocean data assimilation.

\subsection{\acronym training network and loss}

To train the network using partially observed data and evaluate it outside the observed domain, we design a multilayer algorithm, using the fact that the ocean model is represented by vertically stacked layers, each with homogeneous density but differing from one layer to the next in a \av{stable} stratified structure.

In the context of the three-layer quasi-geostrophic model, we consider three differential operators $\D_1, \D_2, \D_3$ allowing us to rewrite quasi-geostrophic dynamics \eqref{qg} as:

\begin{linenomath*}\begin{equation}
\left\{
\begin{aligned}
    & \D_1({\psi}_1, \nabla {\psi}_2) = 0 \\
    & \D_2({\psi}_2, \nabla {\psi}_{1}, \nabla{\psi}_{3}) = 0 \\
    & \D_3({\psi}_3, \nabla {\psi}_{2}) = 0   
\end{aligned}
\right.
\label{qgPDE}
\end{equation}\end{linenomath*}
with $\nabla = (\partial_t, \partial_x, \partial_y)$. We note that each layer interacts only with the layers directly in contact with it, as illustrated in Fig.~\ref{qg_framework}(a). For example, the second layer interacts non-linearly with both the first and third layers. More generally, the operator $\D_\ell$ in layer $\ell$ involves only the stream function $\psi_\ell$ of the current layer and the temporal and spatial derivatives of $\psi_{\ell-1}$ and $\psi_{\ell+1}$.

This peculiar vertical structure of the flow model organized in layers encourages us  to use separated networks to predict the stream functions and let them interact during their learning, instead of relying on a single network to generate direct estimates of $(\psi_1, \psi_2, \psi_3)$.  In what follows, we will reconstruct the three fields of stream functions with three separate networks. We will use the following property: $\psi_\ell$ knows about $\nabla  \psi_{\ell+1}$ and $\nabla  \psi_{\ell-1}$ but not directly $\psi_{\ell+1}$ or $\psi_{\ell-1}$. Consequently, for each stratification layer $\ell$, we design a network $\Psi_{\theta_\ell}$ taking $(t, x, y)$ as input allowing to estimate $\psi_\ell$, with network parameters $\theta_\ell$. Each network will take the form of a SIREN as described in~\eqref{SIREN}.

The proposed minimization problem aiming to estimate the parameters $(\theta_1, \theta_{2}, \theta_{3})$ is formulated as the minimisation of the loss function $\mathscr{L}_{\texttt{\acronym}}$: 
\begin{linenomath*}\begin{align}\begin{split}
    \mathscr{L}_{\texttt{\acronym}}(\theta_1, \theta_2, \theta_3) = & \frac{1}{\vert \mathcal{I}_1 \vert}\sum_{i\in \mathcal{I}_1}|{\Psi}_{\theta_1}(t_i, x_i, y_i) - d_i|^2 + 
    \Vert \D_1({\Psi}_{\theta_1}, \nabla {\Psi}_{\theta_2})\Vert \\ & + \sum_{\ell=2}^3\frac{1}{\vert \mathcal{I}_\ell \vert}\sum_{i\in \mathcal{I}_\ell}|{\Psi}_{\theta_\ell}(t_i, x_i, y_i) - {\Psi}_{\theta_{\ell-1}}(t_i, x_i, y_i) - d_i|^2 \\ & + \Vert\D_2({\Psi}_{\theta_2}, \nabla{\Psi}_{\theta_1}, \nabla  {\Psi}_{\theta_3}))\Vert + \Vert \D_3({\Psi}_{\theta_3}, \nabla{\Psi}_{\theta_2})\Vert.
\label{loss}
\end{split}\end{align}\end{linenomath*}
The quantity of interest are the estimated fields $\widehat{\psi}_\ell  := \Psi_{\widehat{\theta}_\ell}$ with 
$$
(\widehat{\theta}_1, \widehat{\theta}_2, \widehat{\theta}_3) = \underset{(\theta_1, \theta_2, \theta_3)}{\textrm{argmin}}    \mathscr{L}_{\texttt{\acronym}}(\theta_1, \theta_2, \theta_3).
$$

\begin{figure}[htbp!]
\begin{center}
\includegraphics[height=4cm]{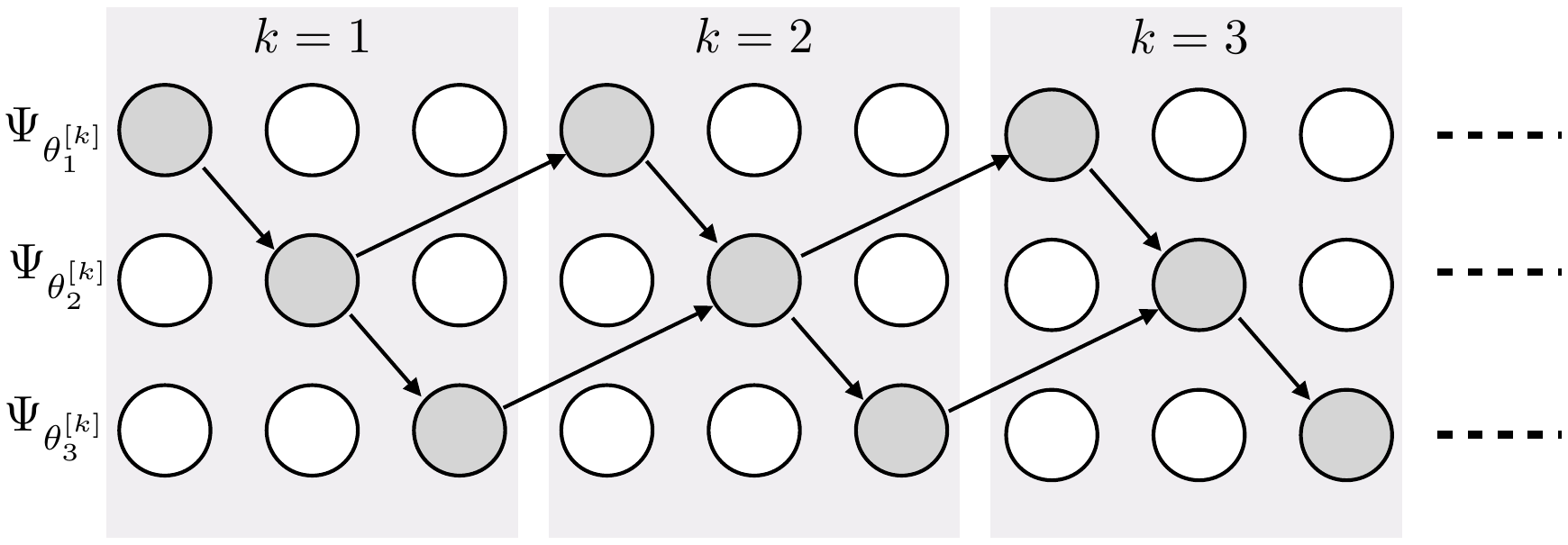}
\end{center}
\caption{Circulation of the information in the proposed multilayer optimisation process. This illustration is provided for the three first iterations for the 3-layer quasi-geostrophic dynamics but can easily be generated to $N$ layers introducing $N$ networks alternatively learning.}
\label{model1}
\end{figure}

In order to look for such a minimiser, we introduce a strategy guided by the stratified structure of the quasi-geostrophic model and the  natural hierarchy between stratification layers. We propose to learn each oceanic layer successively starting from the surface down to the deep layer as sketched in Fig.~\ref{model1} and explained as follows. Such a strategy is a block-minimization procedure that consists in solving, for each iteration $k$,
{\small{\begin{align*}
\theta_1^{[k+1]} &= \underset{\theta_1}{\textrm{argmin}}\; \underbrace{\sum_{i\in \mathcal{I}_1} |\Psi_{\theta_1}(t_i, x_i, y_i) - d_i|^2+ \left\Vert \D_1\left(\Psi_{\theta_1}, \nabla \Psi_{\theta_2^{[k]}}\right)\right\Vert}_{\mathscr{L}^{(1)}_{\texttt{\acronym}}\big(\theta_1, \theta_2^{[k]}\big)}\\
\theta_2^{[k+1]} &= \underset{\theta_2}{\textrm{argmin}}\;  \underbrace{\sum_{i\in \mathcal{I}_2} |{\Psi}_{\theta_2}(t_i, x_i, y_i) - {\Psi}_{\theta_1^{[k+1]}}(t_i, x_i, y_i) - d_i|^2+  \left\Vert \D_2\left(\Psi_{\theta_2}, \nabla \Psi_{\theta_1^{[k+1]}}, \nabla \Psi_{\theta_3^{[k]}}\right)\right\Vert}_{ \mathscr{L}^{(2)}_{\texttt{\acronym}}\big(\theta_1^{[k+1]}, \theta_2,\theta_3^{[k]}\big)} \\
\theta_3^{[k+1]} &= \underset{\theta_3}{\textrm{argmin}}\; \underbrace{\sum_{i\in \mathcal{I}_3} |{\Psi}_{\theta_3}(t_i, x_i, y_i) - {\Psi}_{\theta_2^{[k+1]}}(t_i, x_i, y_i) - d_i|^2+ \left \Vert \D_3\left (\Psi_{\theta_3}, \nabla\Psi_{\theta_2^{[k+1]}}\right )\right\Vert}_{ \mathscr{L}^{(3)}_{\texttt{\acronym}}\big(\theta_2^{[k+1]},\theta_3\big)}\\
\end{align*}}}

Once the training of $\Psi_{\theta_1}$ has converged, we can  compute an estimation of  $\nabla \Psi_{\theta_1}$ that we freeze (meaning we stop learning $\theta_1$) in order to define the approximate differential operator $\D_2(\,\cdot\,, \nabla \Psi_{\theta_1}, \,\cdot\,)$ and start learning the parameters of $\Psi_{\theta_2}$ using the loss $\mathscr{L}^{(2)}_{\texttt{\acronym}}$. Finally, we construct $\D_3$ similarly using a frozen estimate of $\nabla \Psi_2$ to learn the parameters of $\Psi_{\theta_3}$ using the loss $\mathscr{L}^{(3)}_{\texttt{\acronym}}$. The circulation of information is sketched in the diagram Fig.~\ref{model1}, and the algorithm is summed up in Alg. \ref{alg1}.

The algorithm procedure is initialized with $\Psi_{\theta_2^{[0]}} = \Psi_{\theta_3^{[0]}} = 0$. The motivation behind the choice $\Psi_{\theta_2^{[0]}} =  0$ is that the flow is much stronger at the surface than in the interior of the domain, i.e. $\nabla\psi_1 \gg \nabla\psi_2$. Therefore, a fair approximation for $\D_1$ is  to consider $\nabla \psi_2 = 0$ everywhere.  Similarly, the flow in the intermediate layer is much stronger than in the abyss, hence the choice $\Psi_{\theta_3^{[0]}} =  0$.

\begin{algorithm}
\caption{\acronym \label{alg1}}
\begin{algorithmic}
\Require Observations points: $(\mathcal{I}_1,\mathcal{I}_2,\mathcal{I}_3)$ and $\{(t_i, x_i, y_i, d_i)\}_{i\in \mathcal{I}_\ell}$ for $\ell=1,2,3$. \\  
$\qquad\;\;\;\;\;$ Collocation points: $\left\{(T_j, X_j, Y_j)\right\}_{j\in \mathcal{J}}$.

\While{$ \mathscr{L}_{\texttt{\acronym}} \geq \varepsilon$}
\For {k=1 to K}
\State Get batchs from $\mathcal{I}$  and define collocation points
\State $\theta_1^{[k+1]} = \underset{\theta_1}{\argmin} \; \mathscr{L}^{(1)}_{\texttt{\acronym}}\big(\theta_1, \theta_2^{[k]}\big)$
\State $\theta_2^{[k+1]} = \underset{\theta_2}{\argmin} \; \mathscr{L}^{(2)}_{\texttt{\acronym}}\big(\theta_1^{[k+1]}, \theta_2,\theta_3^{[k]}\big)$
\State $\theta_3^{[k+1]} = \underset{\theta_3}{\argmin} \; \mathscr{L}^{(3)}_{\texttt{\acronym}}\big(\theta_2^{[k+1]},\theta_3\big)$
\EndFor
\EndWhile
\end{algorithmic}
\end{algorithm}

\newpage
\section{Numerical experiments}

\subsection{Data}\label{sub:data}

The ground truth $(\psi_1,\psi_2,\psi_3)$ are obtained from the three layers quasi-geostrophic model of the ocean introduced section \ref{MultiQG}. The values implemented in the studied simulation are summarized in Tab.~\ref{tab:coeffs}. The lengthscales of the system are: the size of the domain $L = 4000$~km and the Rossby deformation radii are $L_R\sim \mathcal{O}(30)$~km. The corresponding timescales are of order $T = 1/(\beta L_R)\sim22$~days.

Typical flow fields obtained with this model are shown in Fig.~\ref{qg_framework}. In the upper layer, a western-intensified front is visible at the interface between two large-scale gyres, corresponding to a coherent eastward jet similar to the Gulf Stream or Kuroshio. Coherent eddies can also be seen propagating westward, corresponding to mesoscale oceanic rings. Additionally, Rossby waves propagate across the entire domain, with their phase moving westward. In the interior layers (Fig.~\ref{qg_framework}(b)), the front, eddies, and waves are still present, supporting our hypothesis that surface data can be used to help assimilate ocean interior information. To summarize, this ground truth simulation features simple yet interesting flow structures, both horizontally and vertically, making it an ideal framework for validating our reconstruction method. 

From these simulated data, we constructed pseudo-observations, motivated by the framework presented in section \ref{Pseud-Obs prob}, to combine SWOT-like data and ARGO-like data. The pseudo-observation locations are organized into three sets, $\mathcal{I}_\ell$, each associated with a layer $\ell$. In the  upper layer,  data correspond to SWOT-like observations, providing measurements of $\psi_1$. In the second and bottom layers, data correspond to ARGO-like observations, providing values of $\psi_2 - \psi_1$ and $\psi_3 - \psi_2$, respectively.

A visual representation of the sets $\mathcal{I}_\ell$ is displayed in Fig.~\ref{pseudo-obs}. The following three paragraphs detail how these sets are constructed in the surface, the interior and the bottom layer.

The set $\mathcal{I}_1$ in the surface layer is obtained by constructing a mask that replicates the trace of a virtual SWOT-like satellite observation of sea surface height, following \citeA{leguillou.metref.ea_2021}. To do this, we virtually placed our ocean basin at 40$^o$N. We then used the coordinates of the actual SWOT satellite trajectory and its two swaths, as if it passed over that domain. $\mathcal{I}_1$ is the set of $(t, x, y)$ coordinates for which the point $(x, y)$ belongs to one of the swaths of SWOT at time $t$. Over a period of 100 days, this results in a total of 392,470 observation points, \textit{i.e.}, approximately $1.5\%$ of the domain being observed each day.

The set $\mathcal{I}_2$ in the intermediate layer corresponds to the locations of 500 independent ARGO-like buoys, with daily observations. The buoy locations are fixed for 22 days and then randomly shuffled across the $513 \times 513$ domain grid for the next cycle.

The set $\mathcal{I}_3$ in the deeper layer is obtained by randomly selecting 500 buoy locations in the $513 \times 513$ grid, providing measurements every 10 days. This contrasts with the daily sampling of the intermediate layer. The buoy locations of this layer are also fixed for 22 days and then reshuffled, just as in the intermediate layer. The lower temporal sampling of the deeper layer indirectly accounts for the scarcity of deep ARGO floats.

\subsection{Algorithm}

Before providing a detailed account of our architecture and hyperparameter choices, we must note that PINNs are known to be finicky to train \cite{raissi2020hidden}. Although we operate in a reconstruction regime close to interpolation (where the physics component of the loss acts as a regularization term), we still face three major challenges: spectral bias, loss imbalance, and sensitivity to collocation points. First, the spectral bias is a general failure mode in neural network training, where the network struggles to learn high-frequency components of a signal \cite{rahaman2019spectral}. This issue persists even when the physics-based loss term is disregarded. Among possible solutions, the use of a SIREN architecture with proper initialization helps bypass this problem \cite{sitzmann2020implicit,combette2026new}. Second, the loss imbalance refers to the difficulty of selecting an appropriate regularization parameter, $\lambda$, to balance the data fidelity term and the physics-informed term in the loss function, Eq.~\ref{obj}. Indeed, the physics-based regularization can introduce noise into the loss landscape, slowing down or even preventing convergence during training \cite{rathore2024challenges}. Finally, the choice of collocation points plays a critical role \cite{nguyen2022fixed}. This choice is central to estimating the computational cost of training, as evaluating the physics-informed loss at these points may require up to three full differentiation operations of the network with respect to the input variables (e.g., differentiating the quadratic norm of diffusion terms). Consequently, the memory cost per gradient step increases rapidly with the number of collocation points. In the following section, we detail the choices we made to mitigate these three issues.

We deploy three SIREN neural networks, one for each layer. In our experiments we explore the impact of $\Omega_t, K_x, K_y$, controlling the weights of SIREN's first layer as  they are the only parameters of the network that can directly be physically interpreted. Indeed, had we taken all the multiplicative weights after the first layer (the $W$'s in \eqref{SIREN}) to be small and the biases to be zeroed (the $b$'s), we could rewrite \eqref{SIREN} as: $\Psi_\theta(t, x, y) \approx W\cdot\sin(K_xx+K_yy - \Omega_t +\varphi)$. In this case, the parameters would approach the Fourier transform of the signal. However, we set the biases to be non-zero to allow for more expressivity of the networks. In their paper, \citeA{sitzmann2020implicit} set a maximal value $\omega_0$ for the values of parameters in first layer and initialise the weight uniformly between $-\omega_0$ and $\omega_0$. This could correspond to a maximal allowed wavenumbers to avoid aliasing. Conducting numerical experiments, we noticed in our case that the values in $K_x$, $K_y$ and $\Omega_t$ do not change significantly during learning. We thus decided to set these values by hand in the following way. We first chose a maximal value for the pulsation $\omega_{max}$ and one for the wave numbers $k_{max}$. Then we fill $\Omega_t$ with $N$ uniformly spread numbers between $-\omega_{max}$ and $\omega_{max}$, $K_x$ with $N$ uniformly spread numbers between $-k_{max}$ and $k_{max}$ and finally $K_y$ with $N$ uniformly spread numbers between 0 and $k_{max}$. This way, the first layer covers a wide variety of potential waves propagating in our domain of the form $\sin(k_xx + k_yy - \omega_tt)$ with $k_x \in K_x$, $k_y \in K_y$ and $\omega_t \in \Omega_t$. This choice leads to $N^3$ waves. Choosing $N=5$ is what set the width of our network to equal 125. Note that adding negative values for $k_x$ does not express more waves as the sine function is odd.

Other important parameters to adjust that can affect the training of the network are the number and the location of the collocation points. The higher it is, the better the estimation of $\|\D(\cdot)\|_{L^2(\Omega)}$. However, the number of collocation point to consider is limited by the memory capacity of the machine used. We opted for $\texttt{batch\_size\_col} = 100\,000$ collocation points randomly placed at every gradient step in the network's $[-1, 1]^3$ input domain. Compared to the PINNs literature, this is relatively few but the goal of our PINN-like loss term is to regularize the reconstruction and not numerically solve the equations of motion. It is usually recommended to have two or more orders of magnitude difference between the number of ``boundary condition" points, here observations, and that of collocation points but we had satisfying results with this set-up. Adding more collocation points would significantly increase the memory requirements of the method and probably call for multi-GPU training, which we did not explore. 

We look for the minimiser of the loss function using the gradient-based optimizer ADAM with learning rate of 1e-4 and  $13$ batches per epoch for the surface layer and $10$ for the interior ones. In accordance with other PINNs formulation, the input and output are normalized.
This is detailed in \ref{Norm}. All experiments were executed on a machine with an Intel Xeon Silver 4215R CPU and an NVIDIA A100 GPU with 40 GB of RAM.
The main parameter to tune is the regularisation parameter $\lambda$, the ratio of the data fidelity loss term to the  physic-informed loss term $\mathcal{L}_{dyn}$. It balances these two parts of the objective function and controls how much the network trusts the observations during its learning and how much the reconstruction is a physical field. A high value for $\lambda$ will thus lead to a very smooth reconstruction. We chose $\lambda=0.0005$ following the results shown in section \ref{SWOT_results}.

\noindent The values of the hyper-parameters used in the following numerical experiments are summarized in Tab.~\ref{hyperparams}.

\begin{table}[h]
    \centering
    \begin{tabular}{|c|c||c|c|}
        \hline 
        Parameter & Value & Parameter & Value \\
        \hline
        \texttt{N\_obs\_surf} & 392470 & \texttt{N\_obs\_int} & 50000 \\
        \texttt{N\_obs\_bot} & 5000 & \texttt{N\_col} & 1300000 \\
        \texttt{bs\_obs\_surf} & 30190 & \texttt{bs\_obs\_int} & 5000 \\
        \texttt{bs\_obs\_bot} & 500 & \texttt{bs\_col} & 100000 \\
        $k_{max}$ & 10 & $\omega_{max}$ & 1.5 \\
        $\lambda$ & 5e-4 & learning rate & 1e-4 \\
        \hline
    \end{tabular}
    \caption{Summary of the choices of hyper-parameters values in the network. \av{Parameters \texttt{bs\_obs} and \texttt{bs\_col} respectively stand for the batch sizes of observations and collocation points fed to the network at each gradient descent step.}}
    \label{hyperparams}
\end{table}

\subsection{Results}

\begin{figure}[htbp!]
    \centering
    \begin{subfigure}{\textwidth}
        \includegraphics[width=\textwidth]{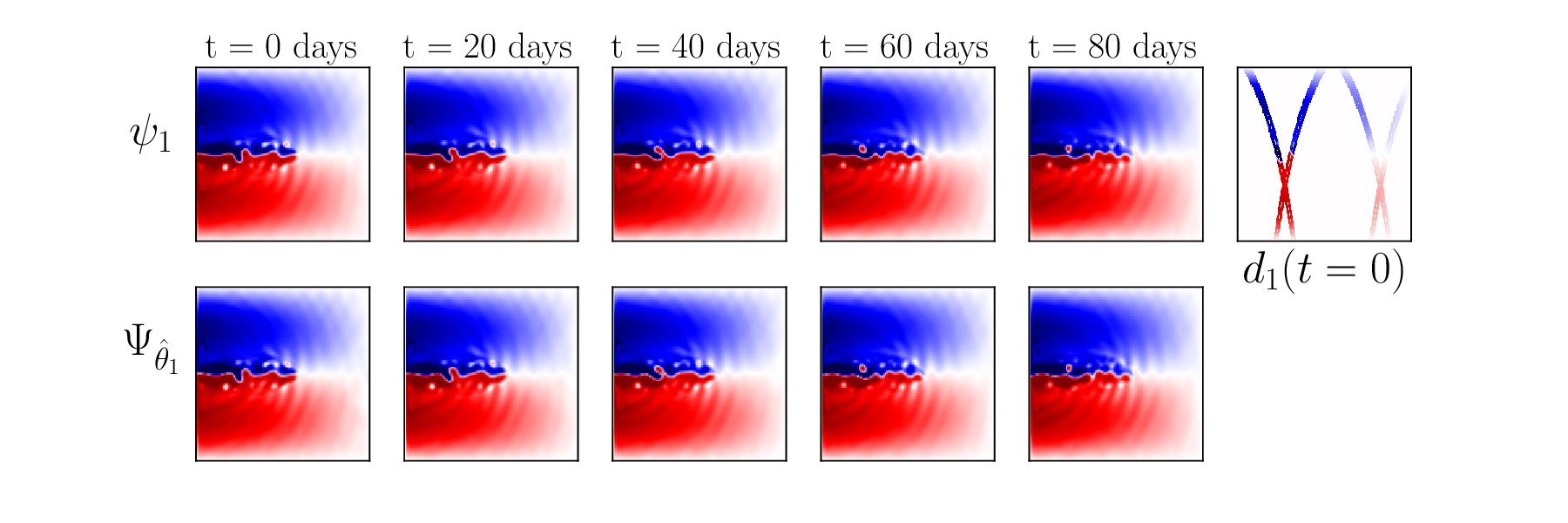}
    \end{subfigure}
    \hfill
    \begin{subfigure}{\textwidth}
        \includegraphics[width=\textwidth]{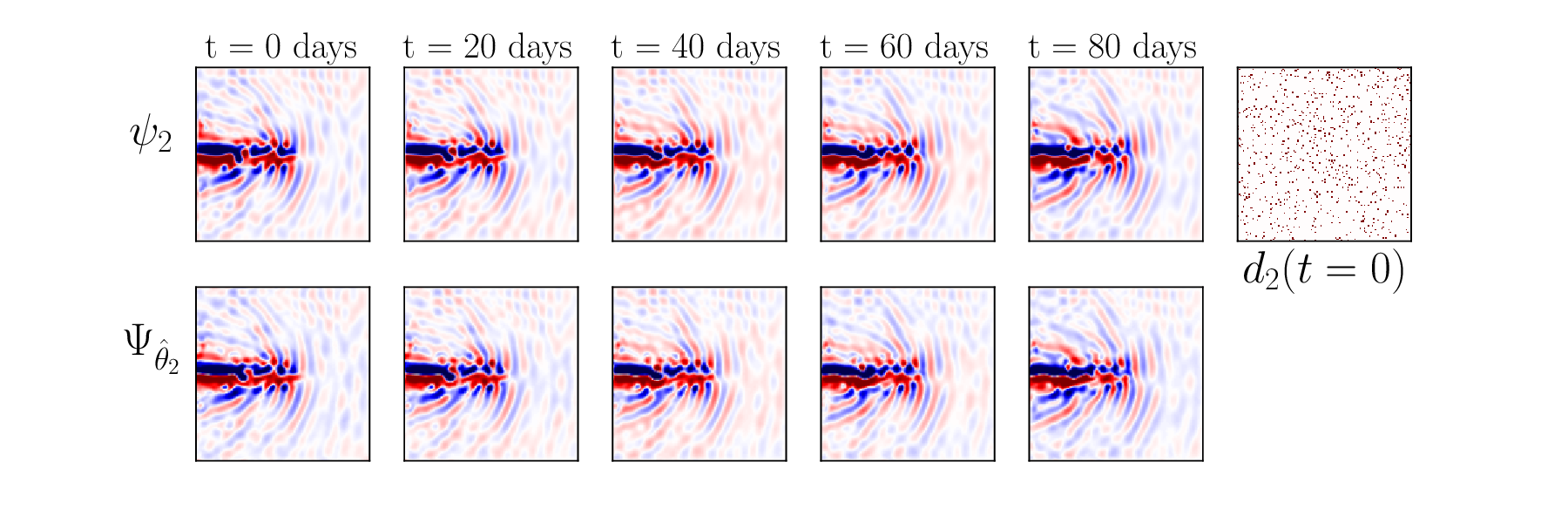}
    \end{subfigure}
    \hfill
    \begin{subfigure}{\textwidth}
        \includegraphics[width=\textwidth]{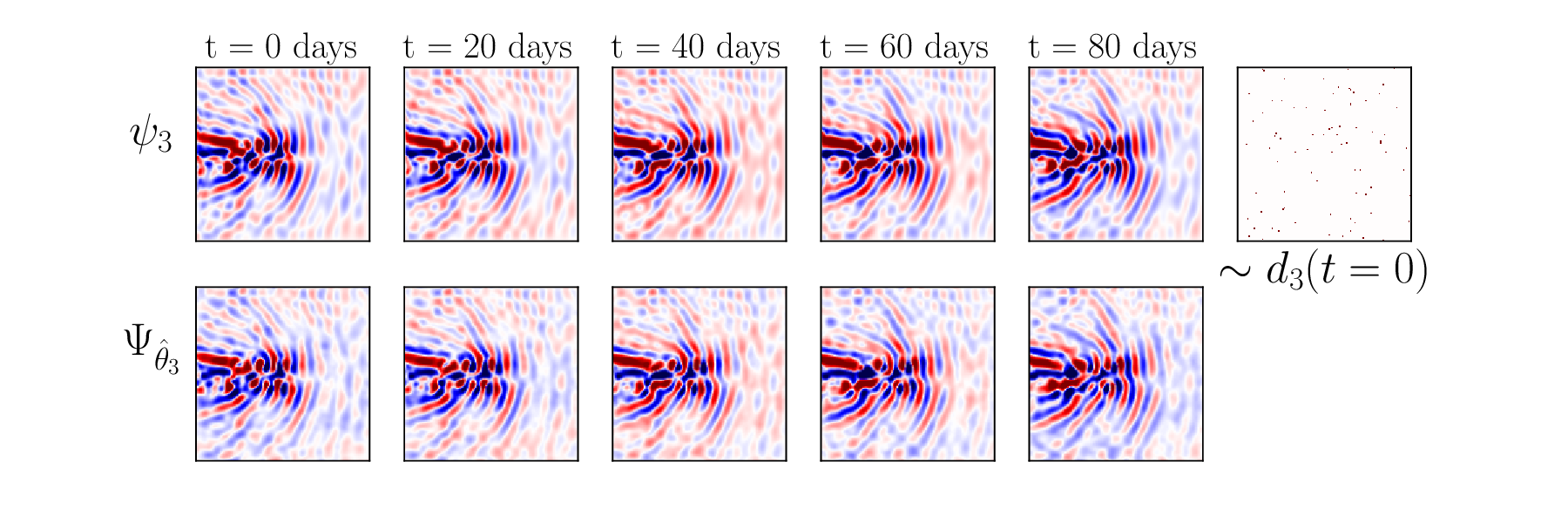}
    \end{subfigure}
    \caption{Reconstructed stream functions in each oceanic layer of the quasi-geostrophic model compared to the full simulated fields. A snapshot of the observation points is shown next to the reconstruction to have an idea of the available data. Note that for the bottom layer, these observations are only available every ten days. \av{The jet and the vortices detaching from it are well resolved even though they are poorly observed. The reconstruction also captures the large scale waves visible mainly in the right part of the domain for the first layer and everywhere for the other two.}}
    \label{SWOT}
\end{figure}

The results presented in Fig.~\ref{SWOT} are obtained after one step of the alternate scheme \ref{method_sketch} on pseudo-observations for a period of 100 days. We were able to reconstruct not only the surface field but also the ones in the interior of the ocean, leveraging physics knowledge to help the information propagate downward. Notably, the reconstructed stream functions include the physical features resolved by the explicit model. Namely, the jet and the vortices detaching from it are well resolved even though they are poorly observed. It also resolved the large scale waves visible mainly in the right part of the domain for the first layer and everywhere for the other two. 
\av{Remarkably, we were able to represent 2d+t stream function fields in each ocean layer, originally resolved at $513 \times 513 \times 100 \approx 3\times 10^7$ grid points, using only $3w + w + 2(w^2 + w) + w + 1\approx  3\times 10^4$ parameters in \acronym. These parameters correspond to the weights and biases of our architecture, which consists of $3$ input neurons and $2$ hidden layers with $w$ neurons each.}

\begin{figure}[htbp!]
    \centering
    \begin{subfigure}{0.5\textwidth}
        \includegraphics[width=\textwidth]{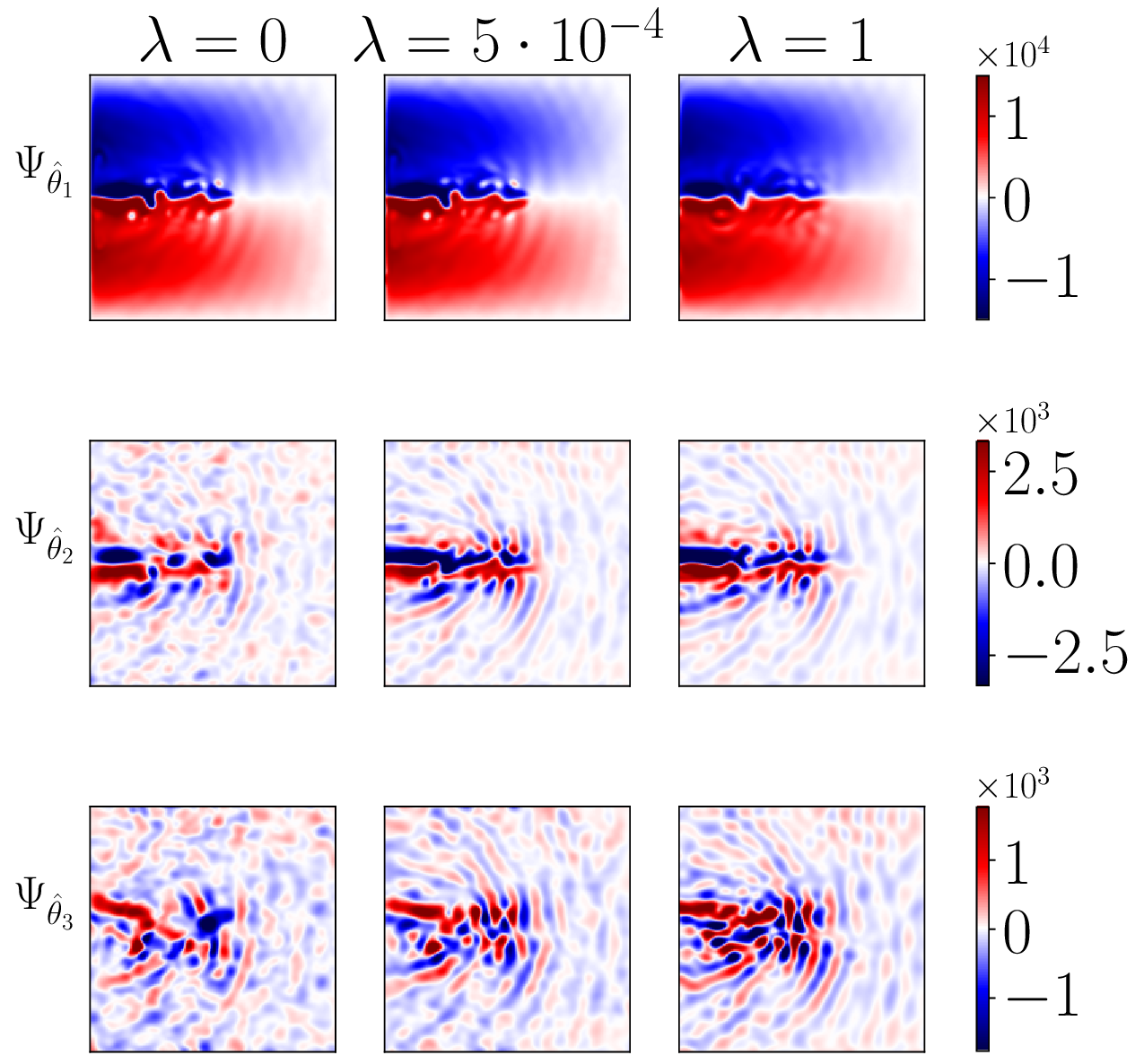}
        \caption{Reconstructed fields.}
    \end{subfigure}
    \hspace{0.1cm}
    \begin{subfigure}{0.45\textwidth}
        \includegraphics[width=\textwidth]{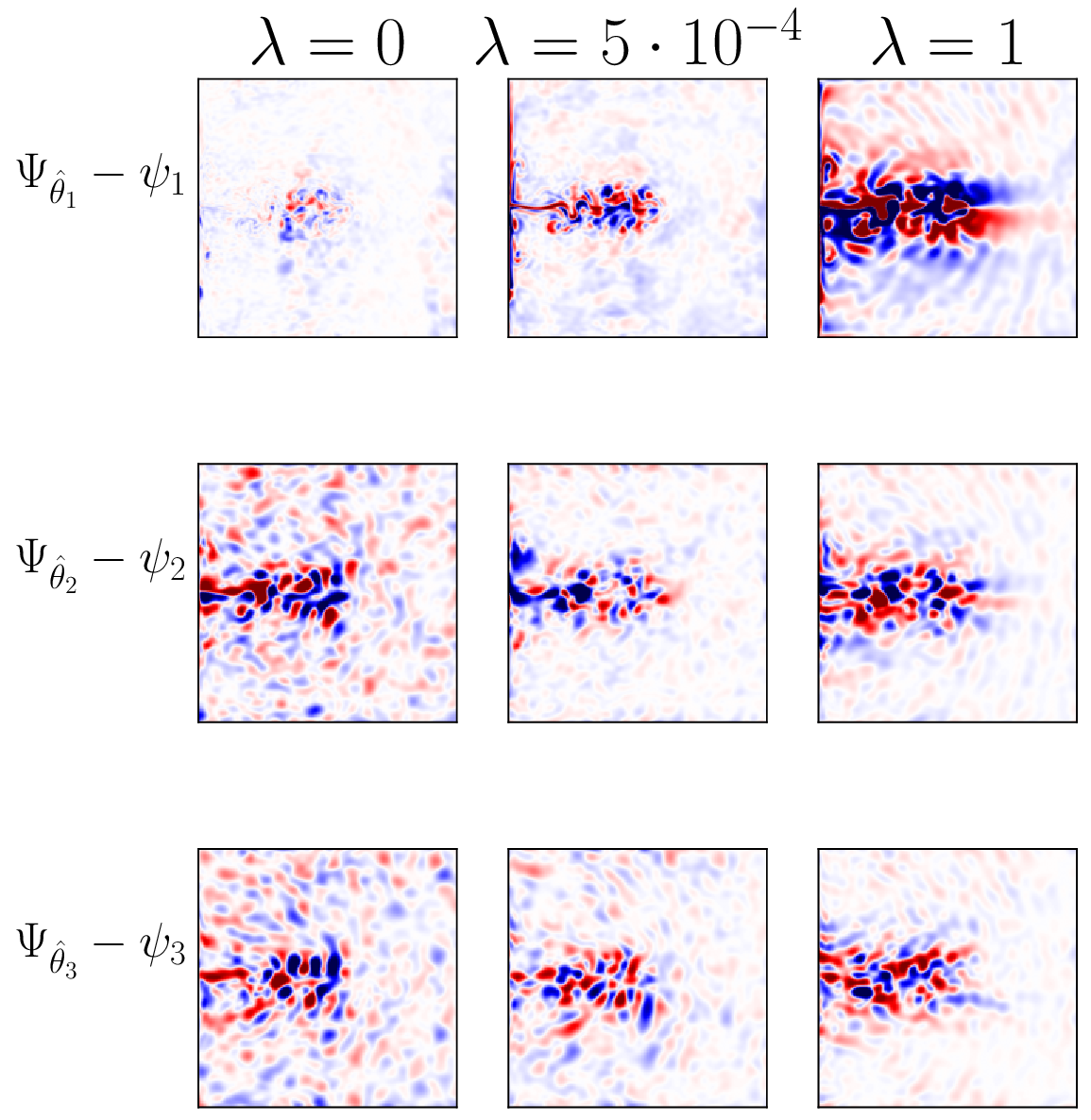}
        \caption{Difference to the ground truth.}
    \end{subfigure}
    \caption{(a) reconstructed fields at a single time step \av{(streamfunction unit is m$^2$.s$^{-1}$)} and (b) the difference to the ground truth simulation for three reconstruction results: no regularisation (i.e. $\lambda=0$), the right amount of regularisation (i.e. $\lambda=10^{-4}$), or too much regularisation (i.e., $\lambda=1$). Qualitatively, one can see from reconstruction error less structure with $\lambda=10^{-4}$ than with $\lambda=0$, showing that the reconstruction captured more faithfully the physical features of the flow.}
    \label{regornoreg}
\end{figure}

\begin{figure}[htbp!]
    \centering
    \begin{subfigure}{0.4\textwidth}
        \includegraphics[width=\textwidth]{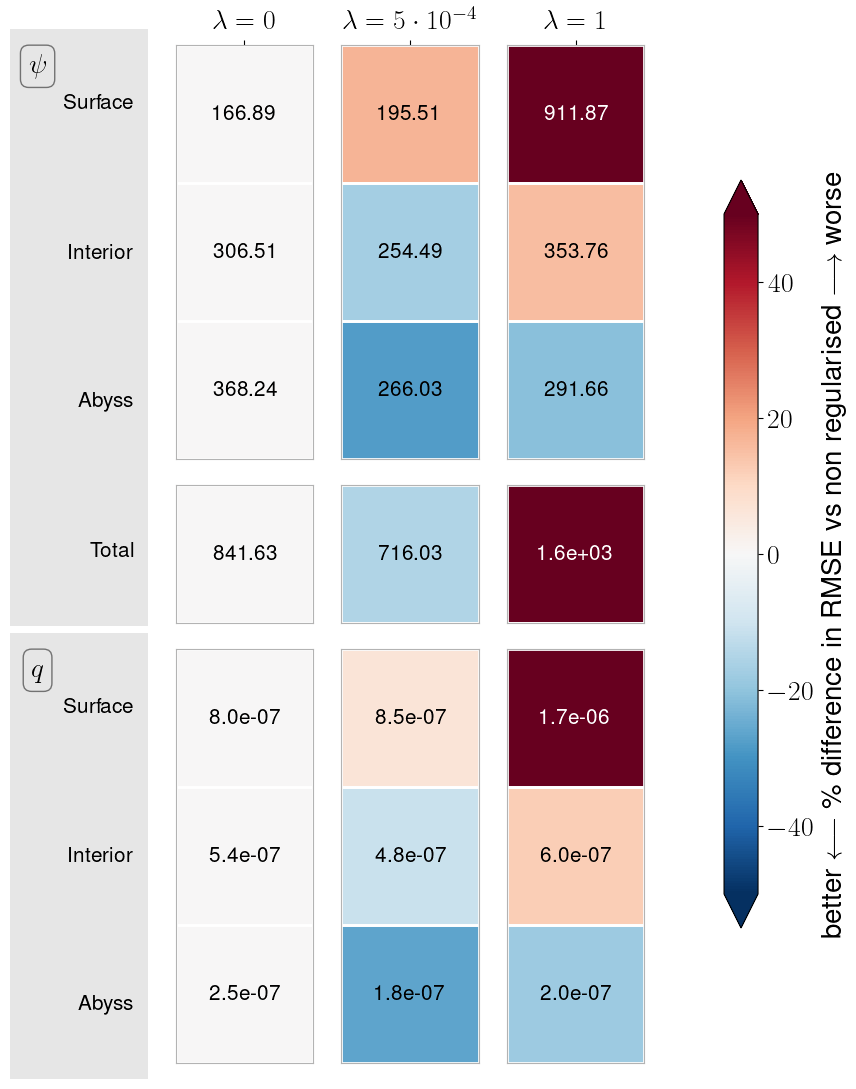}
        \caption{Reconstructed scores.}
        \label{scores}
    \end{subfigure}
    \hspace{0.1cm}
    \begin{subfigure}{0.55\textwidth}
        \includegraphics[width=\textwidth]{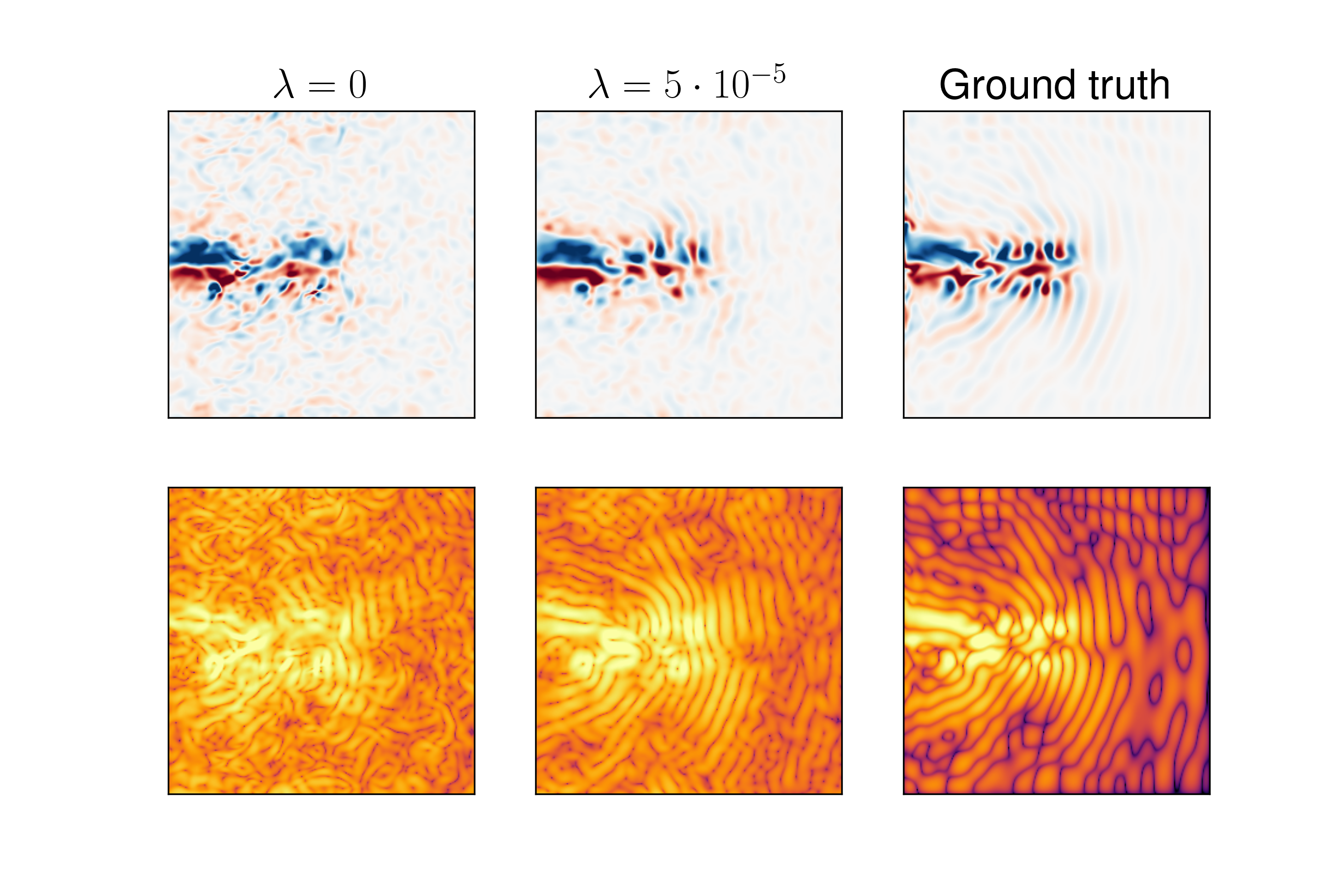}
        \caption{Snapshots of potential vorticity and kinetic energy for non-regularized, regularized and ground truth.}
        \label{physics}
    \end{subfigure}
    \centering
    \caption{\VL{(a) Root mean square error of the reconstruction to the ground truth simulation in $m^2.s^{-1}$ for the stream functions $\psi_{1, 2, 3}$ and $s^{-2}$ for the potential vorticity $q_{1, 2, 3}$. Error in the reconstruction of the deep field was greatly decreased using the physics-informed loss as regularisation with the right balance to the data term. (b) Visual representation of the reconstructed potential vorticity (top row) and kinetic energy (bottom row) in the abyss layer. The reconstructed fields are greatly improved when physical information is used. One can see in both rows how large-scale patterns, corresponding to waves propagating in the abyss, are recovered and were not resolved by the physic-ignorant approach as well as eddies of smaller size.}}
    \label{scoreslam}
\end{figure}

This is an important parameter to tune as it can guide the learning towards a more significant reconstruction even with fewer observations. However, it can yield a degraded reconstruction if wrongly set. The difference in the quality of reconstruction when no regularisation, \textit{i.e.} $\lambda=0$, the right amount of regularisation, $\lambda=5\cdot10^{-4}$ or to much regularisation is applied is shown in Fig.~\ref{regornoreg} at a specific time. Visually, displaying the reconstruction into a small movie shows that \av{structures in the reconstruction field} without regularisation are less smooth than those with regularisation (i.e., jumps in wave phase). Additionally, in the \av{right} column, with increased regularization (i.e., $\lambda=1$), the field becomes smoother, especially for the first layer. The right panel of Fig.~\ref{regornoreg} shows the difference between the reconstructed fields on the left and the ground truth simulation, available everywhere in the domain, \textit{i.e.} the reconstruction error. Qualitatively, one can see less structure in the error field with $\lambda=5\cdot10^{-4}$ than with $\lambda=0$ or $\lambda=1$. This shows the reconstruction captured more faithfully the physical features of the flow. Quantitatively, the mean square value of this difference is lower when some regularisation is used. The reconstruction root mean square error scores evaluated on a $256\times256$ space grid and averaged over the 100 days of assimilation are shown in Fig.~\ref{scores}. At first glance at this figure, it seems that increasing $\lambda$, the weight of physical regularization, affects negatively the reconstruction quality of the surface ocean layers. However it improves greatly the score in the ocean interior layers, and the total reconstruction error is thus decreased. From an optimization point of view this is enough to conclude that our goal to improve the total reconstruction is reached. One would then want to use regularization only in the interior layers but, empirically, a mixed set-up with no regularization other than in the intermediate and abyssal layers is degrading the quality of the interpolation. Fine-tuning the exact procedure is a work left for further investigation. This may show that the physically regularised surface fields have easier information to extract in order to inform the interior fields. In order to verify this, we looked at the reconstruction error of the potential vorticities, defined in Eq.~\eqref{qg}. This is not a target of the training but is an interesting metric for the geophysicists. The result of this test is shown in the lower panel of Fig. \ref{scores} and confirms that these physical quantities are better interpolated when the method is physical constrained which explains why the raw fields' reconstructions are also improved thanks to the multi-level interpolation. A qualitative view of this is given when plotting a snapshot of this potential vorticity and kinetic energy fields of the lowest layer, $q_3$ and $(\partial_x\psi_3^2 + \partial_y\psi_3^2)/2$, reconstructed with and without regularization and comparing it to the ground truth value. This is the content of Fig.\ref{physics}. The reconstructed fields are greatly improved when physical information is used at both large and small scales. We recover large-scale patterns, corresponding to waves propagating in the abyss, that were not resolved by the physic-ignorant approach as well as eddies of smaller size. Plotting the corresponding kinetic energy supports even more this reflection. Plots of snapshots in the other layers are available in \ref{Physics}. One can thus conclude that the ability of SIREN networks to accurately learn a signal and its gradients is remarkable and can be improved thanks to PINNs regularization.

As mentioned above, the initialisation of the layers of SIREN can impact the quality of the reconstructed fields. Exploring the space of hyper-parameters by hand, we found the set of values showed in Tab.~\ref{hyperparams} to work very well and the value of the reconstruction error to be stable around these values.

The number of layers and  neurons per layer were kept low in order to keep a model as transparent as possible for us to choose the aforementioned hyper-parameters by hand.

\label{SWOT_results}
\section{Discussion and conclusions}

We have shown that \acronym is a promising tool for ocean data assimilation by using an idealized three-layer quasi-geostrophic ocean model as the "ground truth" and realistic pseudo-observations resembling surface SWOT-like data and interior ARGO-like data. This minimal setup is relevant for assessing the feasibility of reconstructing deep ocean flows.

The first advantage of \acronym and PINNs in general is that they do not rely on a predefined grid mesh. While there is still an implicit resolution (determined by the number of network parameters), \acronym  can enhance the resolution of the initial field, by filling gaps between observation points, while remaining consistent with the flow dynamics.  A second advantage is that, by contrast to more traditional data assimilation approaches,  there is no need for knowledge of boundary or initial conditions in the loss function. The third advantage of encoding the ocean field in a network is the highly effective representation of the 3d field.
This ability to compress information while maintaining well-behaved derivatives opens up potential for easier manipulation and stokage of those fields.

Several challenges remain to be addressed in future research in order to reconstruct real ocean flows effectively with \acronym, and to quantify uncertainties, thus permitting fair comparison with more traditional data assimilation technics. First, we left aside from our study the central question of uncertainty quantification. This could be tackled with Baysian PINNs or related approaches \cite{yang2021b,wikle2026statistician}. 
 Second, there is significant potential to improve the efficiency of the PINN strategy used in this study. For example, the selection of collocation points used to estimate the physical constraints in the cost function can be optimized, as has been proposed in previous work \cite{Nabian_2021,nguyen2022fixed}. Third, the realism of our pseudo-observation set-up  could be improved: while the number of ARGO floats is realistic for the size of our domain \cite{wong2020argo}, the daily measurement frequency of pseudo-observation is higher than the real temporal sampling, which is closer to 10 days. The choice to reshuffle buoy positions every 22 days is also an idealization, as real buoys would be advected by the flow. In addition, our work has addressed the reconstruction of fields generated by a quasi-geostrophic model, assuming known dynamics and parameters. This represents a necessary first step before tackling more realistic cases, such as the assimilation of 3d+t ocean data, which require richer dynamical models (e.g., primitive equations) and the treatment of unknown parameters. We stress that the idea of layerwise information propagation could be extended naturally to modern isopycnal ocean models, formulated as stacked shallow-water layers.  Alternatively, one may project local 3d data onto a layered quasi-geostrophic model with parameters adjusted at collocation points where physical constraints are enforced. In fact, local quasi-geostrophic simulations have already proven useful in previous studies to interpret and inform global primitive-equation models \cite{venaille2011baroclinic,deremble2023eddy}, and have been used to test other deep flow reconstruction approaches with vision transformers \cite{shan2026testing}. The careful design of projections, following seminal work such as  \cite{flierl1978models}, will be essential in extending our method to more realistic settings, while exploiting the PINN framework to optimize parameter selection.

In summary, this paper demonstrates the potential of Physics-Informed Neural Networks for reconstructing three-dimensional ocean currents from abundant surface data and sparse interior measurements. The novelty of our method lies in its ability to reconstruct not only surface but also deep ocean currents in this context. By combining available data with the underlying dynamics of ocean circulation, PINNs offer a promising approach for inferring deep ocean dynamics. Our study serves as a proof of concept for applying PINNs to deep ocean flow reconstruction.

\newpage
\appendix

\section{Multilayer \acronym~ in the general setting}
\label{appen-sec:general}

In the proposed model, we consider $L$ differential operators $(\D_\ell)_{1,\ldots, L}$ allowing us to rewrite quasi-geostrophic dynamics \eqref{qg} as:

\begin{linenomath*}\begin{equation}
\left\{
\begin{aligned}
    & \D_1(\psi_1, \nabla \psi_2) = 0 \\
    & \D_\ell(\psi_\ell, \nabla \psi_{\ell-1}, \nabla \psi_{\ell+1}) = 0\qquad \forall \ell=2,\ldots,L-1 \\
    & \D_L(\psi_L, \nabla \psi_{L-1}) = 0   
\end{aligned}
\right.
\label{multiqgPDE}
\end{equation}\end{linenomath*}
The optimisation problem we now have to face is then to find a set $(\theta_1, \ldots, \theta_{L-1})$ of parameters that minimise the training loss function $\mathscr{L}_{\texttt{\acronym}}$:

\begin{linenomath*}\begin{align}\begin{split}
    \mathscr{L}_{\texttt{\acronym}}(\theta_1, \theta_2, \ldots, \theta_L) = & \frac{1}{\vert \mathcal{I}_1 \vert}\sum_{i\in \mathcal{I}_1}|{\Psi}_{\theta_1}(t_i, x_i, y_i) - d_i|^2 + 
    \Vert \D_1({\Psi}_{\theta_1}, \nabla {\Psi}_{\theta_2})\Vert \\ + & \sum_{\ell=2}^L\frac{1}{\vert \mathcal{I}_\ell \vert}\sum_{i\in \mathcal{I}_\ell}|{\Psi}_{\theta_\ell}(t_i, x_i, y_i) - {\Psi}_{\theta_{\ell-1}}(t_i, x_i, y_i) - d_i|^2 \\ + & \sum_{\ell=1}^{L-1}\Vert\D_\ell(\Psi_{\theta_\ell}, \nabla \Psi_{\theta_{\ell-1}}, \nabla \Psi_{\theta_{\ell+1}}))\Vert  + \Vert \D_L(\Psi_{\theta_L}, \nabla \Psi_{\theta_{L-1}})\Vert
\label{loss}
\end{split}\end{align}\end{linenomath*}

In order to look for such a minimiser, we introduce a strategy guided by the stratified structure of the quasi-geostrophic model and the  natural hierarchy between stratification layers. We propose to learn each oceanic layer successively starting from the surface down as sketched in Fig.~\ref{model1} and explained as follows. Such a strategy is a block-minimization procedure that consists in solving:

{\small{\begin{align*}
\min_{\theta_1} &\sum_{i\in \mathcal{I}_1} |{\Psi}_{\theta_1}(t_i, x_i, y_i, z_i) - d_i|^2+ \Vert \D_1(\Psi_{\theta_1}, \nabla\Psi_{\theta_2})\Vert\\
\min_{\theta_2} &\sum_{i\in \mathcal{I}_2} |{\Psi}_{\theta_2}(t_i, x_i, y_i) - {\Psi}_{\theta_1}(t_i, x_i, y_i) - d_i|^2+  \Vert \D_2(\Psi_{\theta_2}, \nabla \Psi_{\theta_1}, \nabla \Psi_{\theta_3}))\Vert \\
\vdots\\
\min_{\theta_{L-1}} &\sum_{i\in \mathcal{I}_{L-1}} |{\Psi}_{\theta_{L_1}}(t_i, x_i, y_i) - {\Psi}_{\theta_{L-2}}(t_i, x_i, y_i) - d_i|^2+ \Vert \D_{L-1}(\Psi_{\theta_{L-1}}, \nabla \Psi_{\theta_{L-2}}, \nabla \Psi_{\theta_{L}}))\Vert \\
\min_{\theta_L} &\sum_{i\in \mathcal{I}_L} |{\Psi}_{\theta_L}(t_i, x_i, y_i) - {\Psi}_{\theta_{L-1}}(t_i, x_i, y_i) - d_i|^2+\Vert \D_L(\Psi_{\theta_{L}}, \nabla\Psi_{\theta_{L-1}})\Vert\\
\end{align*}}}

\section{Snapshots of physical quantities}
\label{Physics}

\begin{figure}[htbp!]
    \centering
    \begin{subfigure}{\textwidth}
        \includegraphics[width=0.75\textwidth]{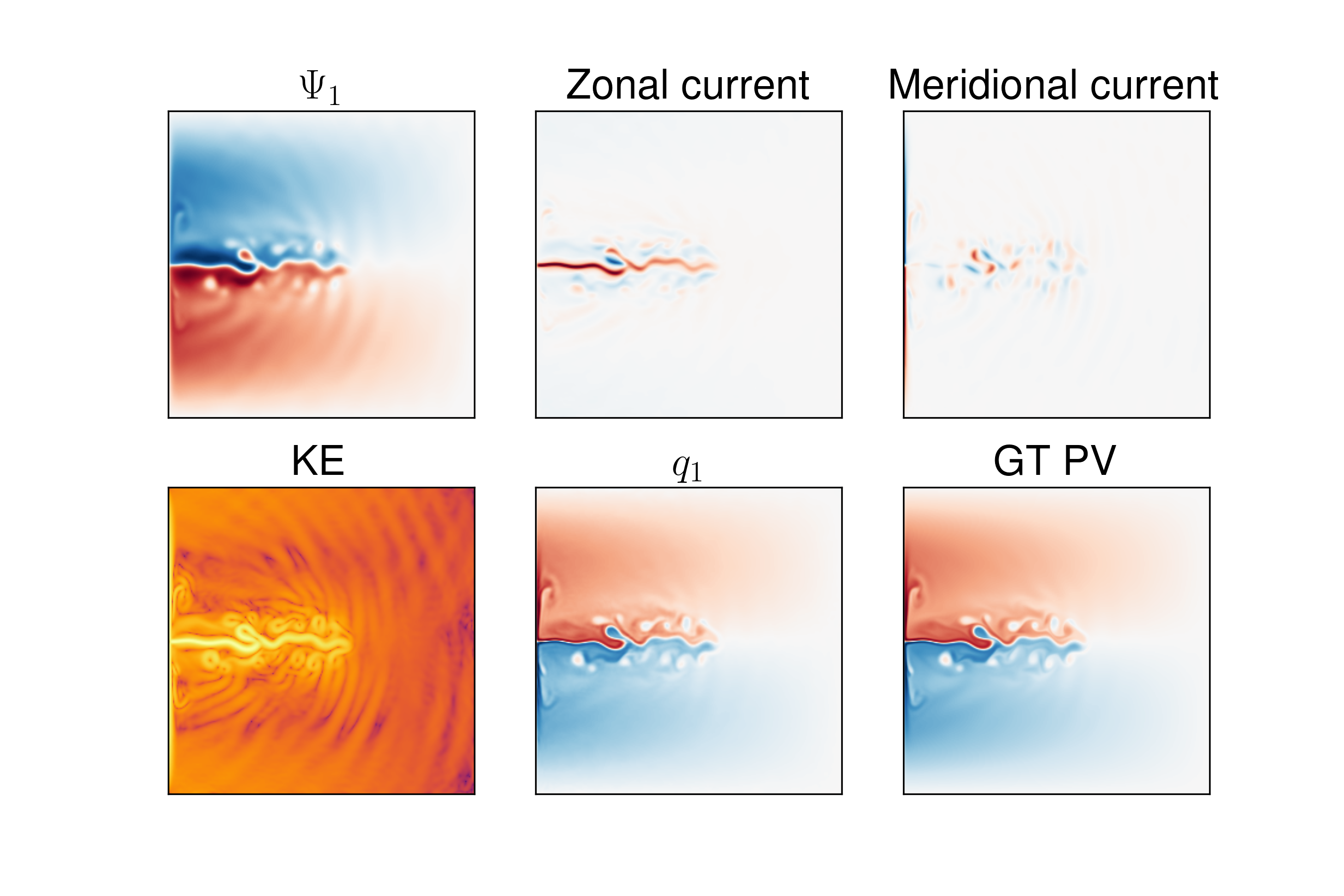}
    \end{subfigure}
    \begin{subfigure}{\textwidth}
        \includegraphics[width=0.75\textwidth]{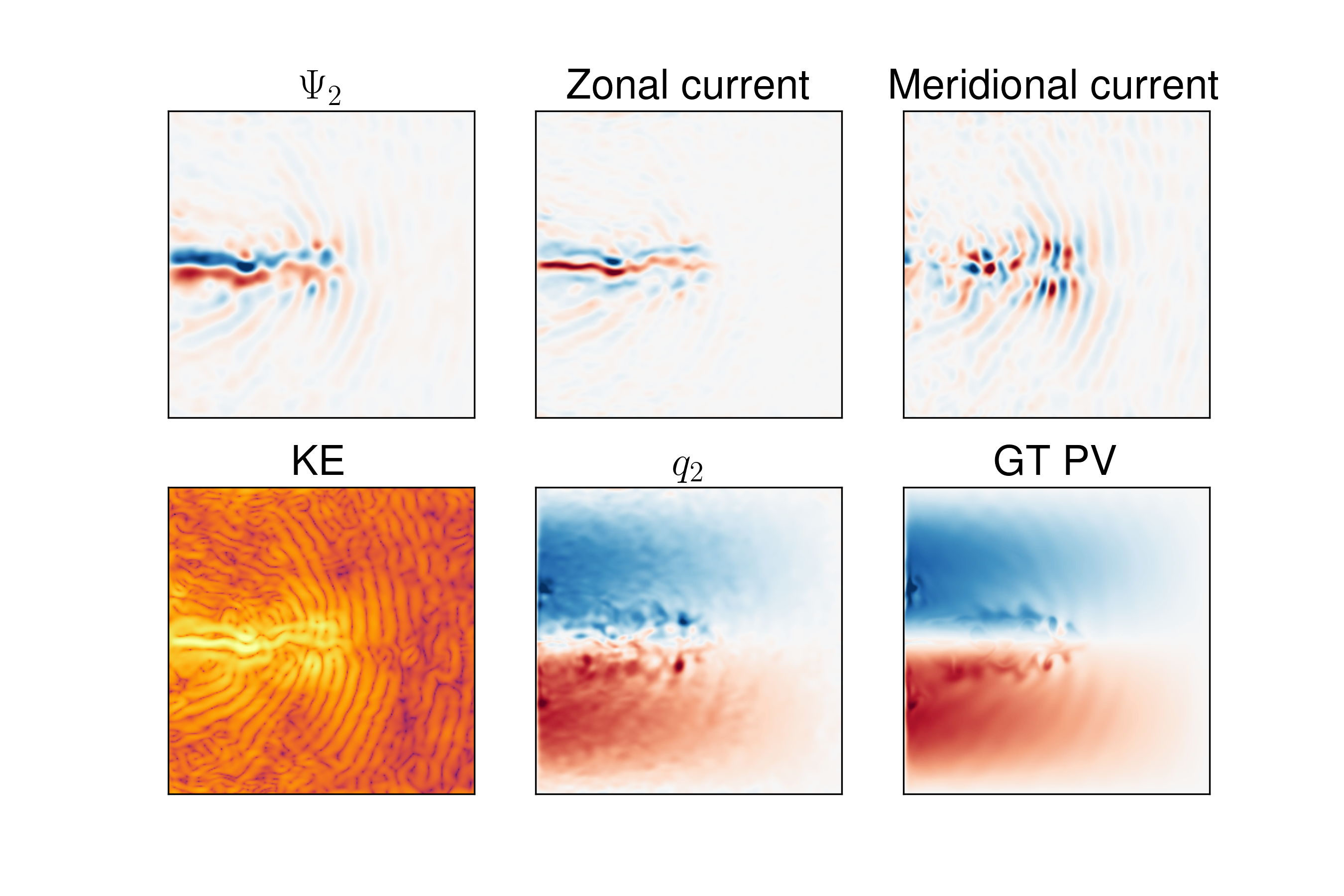}
    \end{subfigure}
    \begin{subfigure}{\textwidth}
        \includegraphics[width=0.75\textwidth]{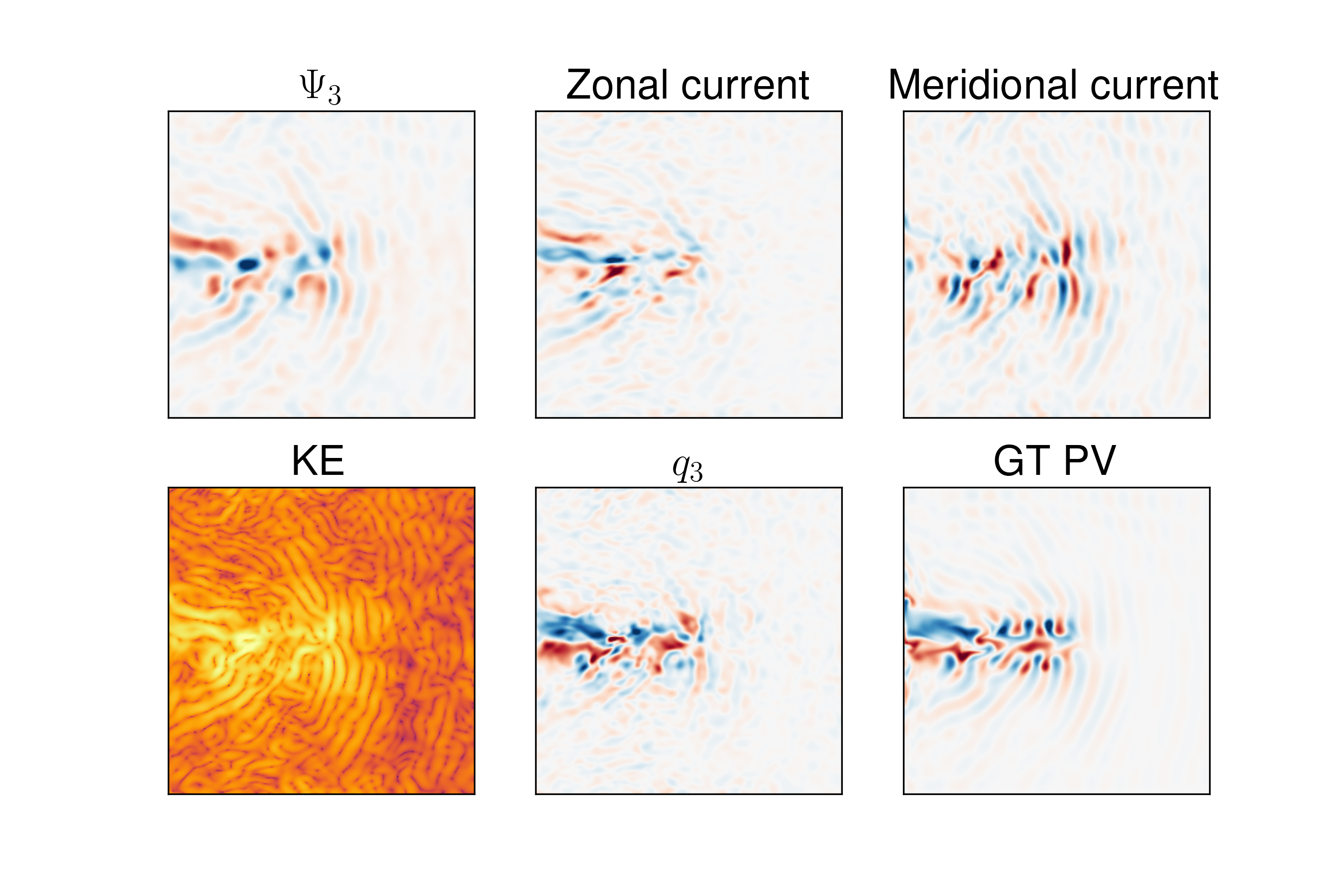}
    \end{subfigure}
    \caption{Snapshots of physical quantities in each ocean layer without physics-informed regularization.}
    \label{reg}
\end{figure}

\begin{figure}[htbp!]
    \centering
    \begin{subfigure}{\textwidth}
        \includegraphics[width=0.75\textwidth]{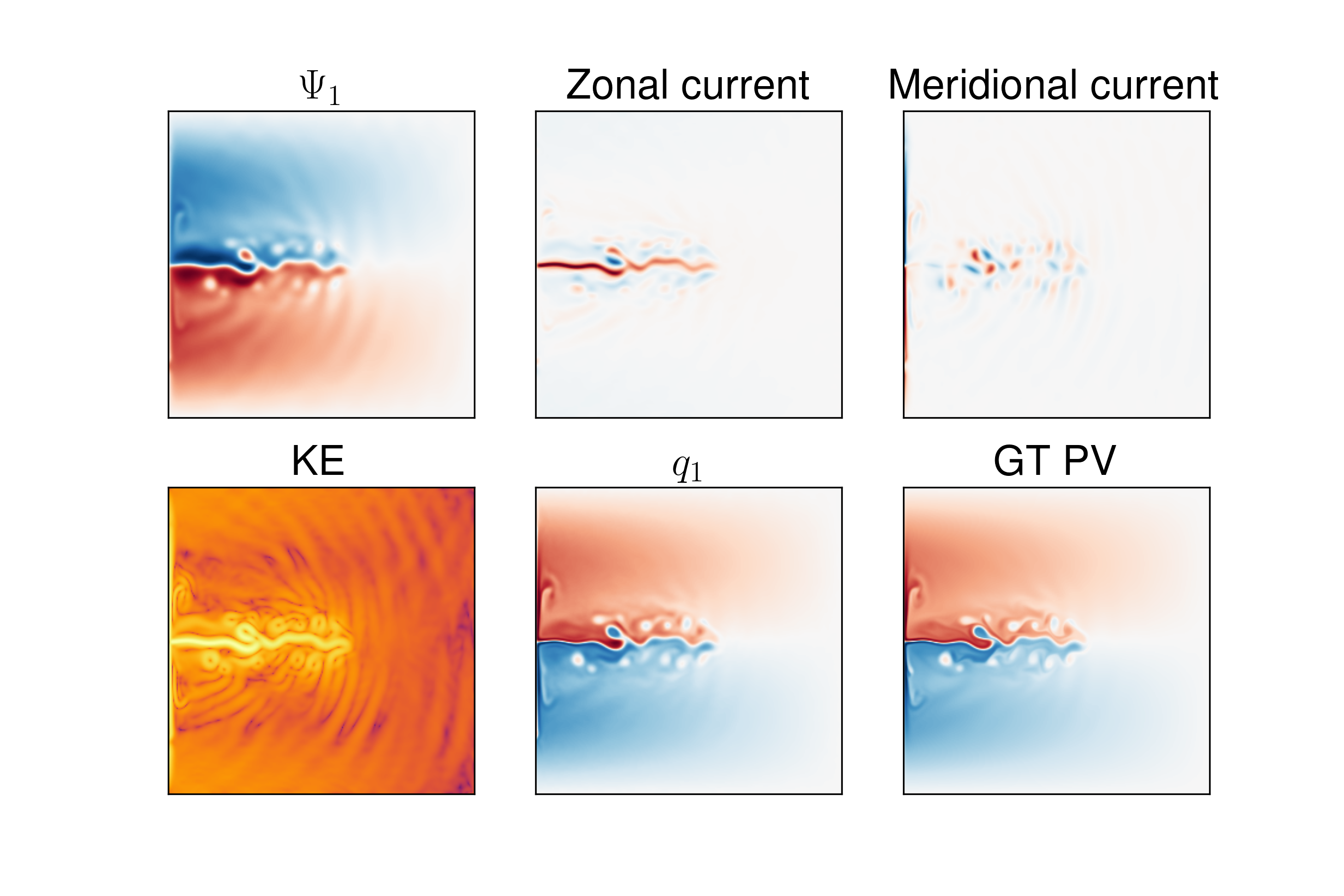}
    \end{subfigure}
    \begin{subfigure}{\textwidth}
        \includegraphics[width=0.75\textwidth]{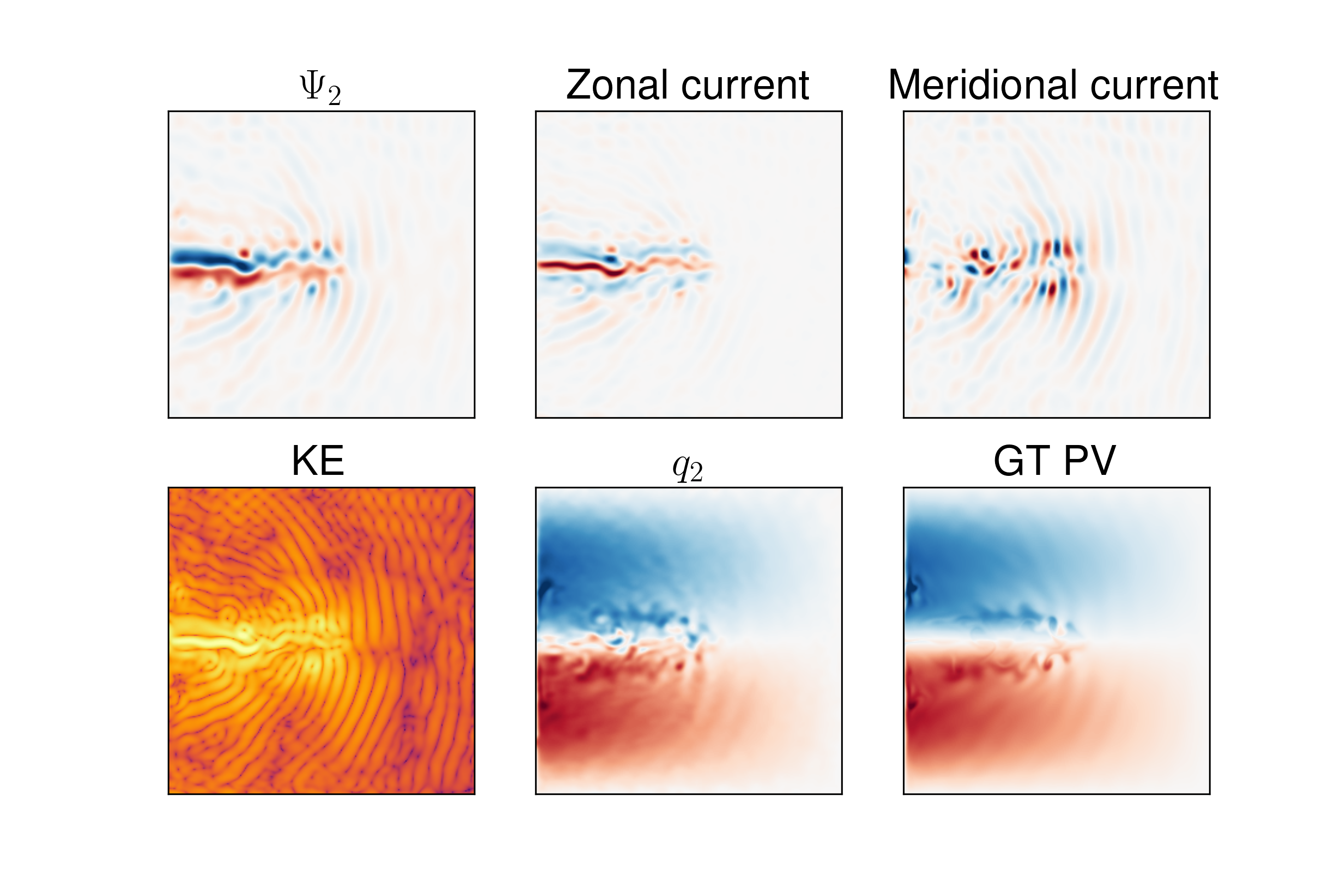}
    \end{subfigure}
    \begin{subfigure}{\textwidth}
        \includegraphics[width=0.75\textwidth]{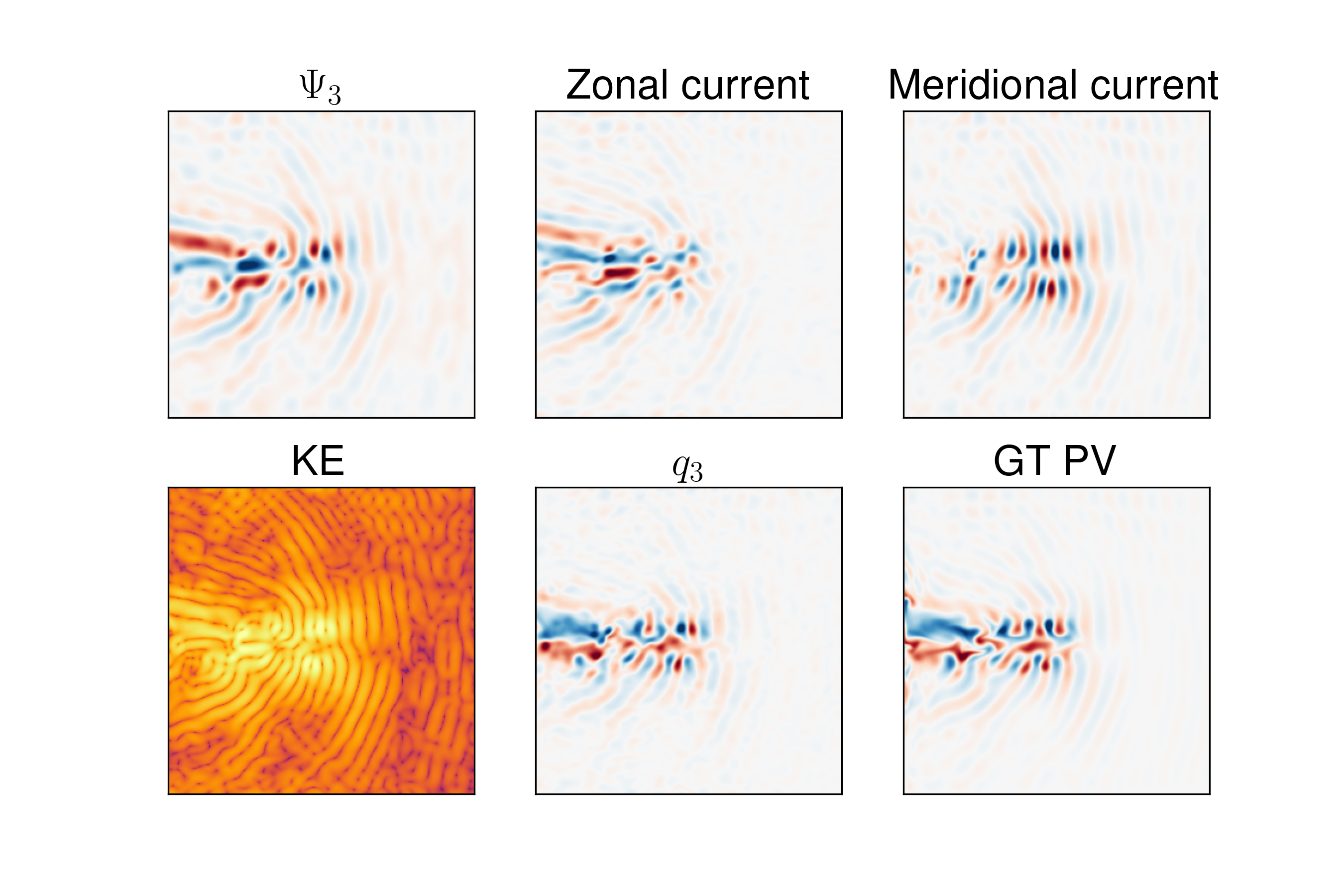}
    \end{subfigure}
    \caption{Snapshots of physical quantities in each ocean layer with physics-informed regularization.}
    \label{noreg}
\end{figure}

Supplementary plots showing the streamfunction, meridional and zonal currents, potential vorticity (PV), kinetic energy (KE) reconstructed with (Fig.\ref{reg}) and without (Fig.\ref{noreg}) regularization. The only physical ground truth available from the reference model other than the streamfunction is the ground truth PV displayed here.

\section{Data normalisation}
\label{Norm}

It is a widely spread information that neural networks tend to prefer normalised inputs and outputs. In order to fit with this idea, we transform the domain from $\Omega=[0, T]\times[0, L]^2$ to $\Tilde{\Omega}=[-1, 1]^3$ and the outputs are normalised as well:

\begin{linenomath*}\begin{equation}
\begin{aligned}
    \psi_1 \longleftarrow (\psi_1 - \mu)/\sigma \\
    \psi_2 \longleftarrow (\psi_2 - \mu)/\sigma \\
    \psi_3 \longleftarrow (\psi_3 - \mu)/\sigma
\end{aligned}
\end{equation}\end{linenomath*}
with $\mu$ and $\sigma$ the mean and standard deviation of the observed data for the first layer.

Now we need to rewrite the transformed differential operators. The advection operators now read:

\begin{linenomath*}\begin{equation}
    \left.\frac{d}{dt}\right|_i = r_t\partial_{\Tilde{t}} + r_x^2 \left[ -\partial_{\Tilde{y}}\left(\sigma\Psi_i\right)\cdot\partial_{\Tilde{x}} + \partial_{\Tilde{x}}\left(\sigma\Psi_i\right)\cdot\partial_{\Tilde{y}} \right]
\end{equation}\end{linenomath*}

Let us write things explicitly for $\D_1$. For given networks $\Psi_1$ and $\Psi_2$ encoding the stream functions in the two first layers, the corresponding encoded potential vorticity $\Tilde{Q}_1$ is:

\begin{linenomath*}\begin{equation}
    \Tilde{Q}_1 = r_x^2\Delta\left(\sigma\Psi_1\right) + \frac{\sigma\Psi_2 - \sigma\Psi_1}{L_1^2} + \beta\frac{\Tilde{y}}{r_x}
\end{equation}\end{linenomath*}

with $r_t = 2/T$ and $r_x = 2/L$. And the operator now reads:

\begin{linenomath*}\begin{equation}
\begin{aligned}
    \D_1(\Psi_1, \nabla\Tilde{\Psi_2}, \Delta\Tilde{\Psi_2}) = & \left.\frac{d}{dt}\right|_1\left(r_x^2\Delta\left(\sigma\Psi_1\right) - \frac{\sigma\Psi_1}{L_1^2}\right)\\
    & + \frac{\sigma}{L_1^2}\left.\frac{d}{dt}\right|_1 \Psi_2 \\
    & + r_x\beta\partial_{\Tilde{x}}\left(\sigma\Psi_1\right) \\
    & - r_x\frac{\tau_0}{H_1}\pi\sin(\pi \Tilde{y}) - \nu r_x^2\Delta \Tilde{Q}_1
\end{aligned}
\end{equation}\end{linenomath*}

The same treatment was done to $\D_2$ and $\D_3$.

\newpage

\section*{Open Research Section}
\noindent The codes implementing \acronym are published in the Zenodo repository 

https://zenodo.org/records/19353310
 
and the idealized ocean simulation data sets (including pseudo-observations) are published in the  Zenodo repository 

https://zenodo.org/records/15074965.

\acknowledgments
  The authors also thank the CBPSMN platform at ENS-Lyon for providing efficient computing resources for all the necessary numerical experiments. The platform uses SIDUS \cite{quemener2013}, which was developed by Emmanuel Quemener \av{and Marianne Corvellec. BD warmly thanks Emmanuel Cosme for useful discussions on deep flow reconstruction in data assimilation. This project was funded by Tremplin CNRS programm, a 80 prime MITI CNRS fellowhip, an received funding from the French National Research Agency (ANR) under the PINOT project, grant agreement
ANR-25-CE56-2607.}.

\bibliography{article}

%
%
%
%
%

\end{document}